%% file: waiacmogeu-cleaned.tex
\RequirePackage[l2tabu, orthodox]{nag}
\RequirePackage{snapshot}

\documentclass[10pt,onecolumn]{article}

\makeatletter
\if@twocolumn
  \usepackage[dvips,letterpaper,top=0.5in, bottom=0.5in, left=0.75in, right=0.5in,includefoot,heightrounded]{geometry}
\else
  \usepackage[dvips,letterpaper,margin=1in,includefoot,heightrounded]{geometry}
\fi

\usepackage{srcltx}

\usepackage{amsmath}
\usepackage{amssymb,amsfonts}

\usepackage{abstract}

\usepackage{graphicx}
\usepackage[usenames,dvipsnames]{color}
\usepackage[usenames,dvipsnames]{xcolor}
\usepackage{subfigure}

\usepackage{booktabs}

\usepackage{setspace}
\usepackage{flushend}
\usepackage{multicol}

\usepackage{cite}
\usepackage{url}
\usepackage[normalem]{ulem}

\usepackage{enumerate}

\usepackage{hyperref}

\usepackage[sc,tiny]{titlesec}

\newcommand{\prd}{\mbox{PRD}}
\newcommand{\CR}{\mbox{CR}}

\newcommand{\mr}{\midrule}
\newcommand{\br}{\bottomrule}

\usepackage{microtype}



\title{%
Wavelet Analysis in a Canine Model of\\
Gastric Electrical Uncoupling
}

\author{%
R.~J.~Cintra%
\thanks{
R. J. Cintra was with Graduate Program in Electrical Engineering,
Universidade Federal de Pernambuco, Brazil.
Currently,
he is with the
\emph{Departamento de Estat\'{\i}stica}
at the same university.
Email: rjdsc@de.ufpe.br}
\and
I.~V.~Tchervensky%
\thanks{
I. V. Tchevensky was
with
the Department of Electrical and Computer Engineering, University of Calgary, Calgary, Alberta, Canada T2N~1N4.
}
\and
V.~S.~Dimitrov%
\thanks{
V. S. Dimitrov
is
with
the Department of Electrical and Computer Engineering, University of Calgary, Calgary, Alberta, Canada T2N~1N4,
and with
the Computer Modelling Group, Calgary.}
\and
M.~P.~Mintchev%
\thanks{
M. P. Mintchev
is
with
the Department of Electrical and Computer Engineering, University of Calgary, Calgary, Alberta, Canada T2N~1N4
and
the
Department of Surgery, University of Alberta, Edmonton, Alberta, Canada T6G~2B7.
Email: mintchev@ucalgary.ca}
}

\date{}

\newcommand{\myabstract}{%
Abnormal gastric motility function could be related to gastric electrical uncoupling,
the lack of electrical, and respectively mechanical, synchronization
in different regions of the stomach.
Therefore, non-invasive detection of the onset of
gastric electrical uncoupling can be important for diagnosing
associated gastric motility disorders.
The aim of this study is to provide a wavelet-based analysis
of electrogastrograms~(EGG, the cutaneous recordings of gastric electric activity),
to detect gastric electric uncoupling.
Eight-channel EGG recordings were acquired from sixteen dogs in
basal state and after each of two circular gastric myotomies.
These myotomies simulated
mild and severe gastric electrical uncoupling,
while keeping the separated gastric sections
electrophysiologically active by preserving their
blood supply.
After visual
inspection, manually selected 10-minute EGG segments were
submitted to wavelet analysis.
Quantitative methodology to choose an optimal wavelet was derived.
This ``matching'' wavelet was determined using the Pollen
parameterization for 6-tap wavelet filters and error minimization
criteria.
After a wavelet-based compression, the distortion of the approximated EGG
signals was computed.
Statistical analysis on the distortion values allowed to
significantly ($p<0.05$) distinguish
basal state from mild and severe gastric electrical uncoupling groups in particular EGG channels.
}

\newcommand{\mykeywords}{%
Electrogastrography, wavelet analysis, signal classification
}

\begin{document}

\makeatletter
\if@twocolumn

\twocolumn[%
  \maketitle
  \begin{onecolabstract}
    \myabstract
  \end{onecolabstract}
  \begin{center}
    \small
    \textbf{Keywords}
    \\\medskip
    \mykeywords
  \end{center}
  \bigskip
]
\saythanks

\else

  \maketitle
  \begin{abstract}
    \myabstract
  \end{abstract}
  \begin{center}
    \small
    \textbf{Keywords}
    \\\medskip
    \mykeywords
  \end{center}
  \bigskip
  \onehalfspacing
\fi

\section{Introduction}

Gastric
electrical activity (GEA) controls the motility of the
stomach~\cite{Szurszewski1981,Daniel94}.
Gastric motility disorders, including functional gastroparesis and
dyspepsia, have been related to alterations in GEA
dynamics~\cite{Borto98}.
Abnormal GEA could be regarded as a
result of two different phenomena:
(i) global gastric electrical dysrhythmias encompassing
simultaneously the entire organ;
and
(ii) local gastric electrical
dysrhythmias, often manifesting themselves as gastric electrical
uncoupling.

Many studies emphasized the impact of alterations in the GEA
rhythm as the main reason for gastric motility
abnormalities~\cite{Chen95,Chen93,Parkman:03}.
However,
global gastric dysrhythmias are by themselves uncommon
and have been objectively registered only incidentally~\cite{YouChey1984,Mintchev97}.
Often, local dysrhythmias and gastric electrical uncoupling are simultaneously present
and could be considered very similar, if not the same phenomena.

Gastric electrical uncoupling occurs when different parts of the stomach lose
synchronization
creating independent oscillating regions
which are dysrhythmic with respect to the global GEA frequency~\cite{Bradshaw03}.
In pure uncoupling these independent regions are characterized
each with steady but different GEA rhythm.
However, often this is not the case, and dysrhythmias are also present.
Regardless whether the stomach is uncoupled or both uncoupled and locally dysrhythmic,
the lack of oscillatory coordination leads to abnormal gastric motor function~\cite{Publi89}.
Therefore, the identification of gastric electrical uncoupling can be
important in diagnosing gastric motility abnormalities~\cite{Daniel94}.

In this context, cutaneous recordings of GEA, known as
electrogastrography (EGG),
 can play a major role in the  diagnosis
of gastric motility disorders~\cite{Smout80}, since EGG was
successfully related to gastric electrical
uncoupling~\cite{Mintchev:97}. Because of its low-cost and
non-invasiveness, the EGG technique has a great appeal as a
clinical tool. Numerous studies had been conducted in order to
classify EGG recordings of surgically produced gastric electrical
uncoupling in large experimental animals. Such studies included
examination of the level of randomness~\cite{Mintchev:1998}, the
level of chaos~\cite{Carre:2001}, biomagnetic field
patterns~\cite{Bradshaw99,Bradshaw03,Bradshaw2003}, and dominant
frequency dynamics~\cite{Mintchev:97}.

Separately, refined signal processing techniques, such as
wavelets, have been employed to analyze
electrogastrograms~\cite{Xie98,Ryu:02,Liang96,Liang02,Qiao96}.
This approach has been used to
(i) propose new wavelets that can offer a better time-frequency localization of
EGG signals~\cite{Xie98,Ryu:02};
(ii) perform noise detection in EGG recordings~\cite{Liang96};
(iii) reduce stimulus artefacts~\cite{Liang02};
and
(iv) characterize global gastric electrical dysrhythmias~\cite{Qiao96}.

In the present study, a new application of wavelets for EGG analysis
in detecting gastric electrical uncoupling and local dysrhythmias is proposed.
It is hypothesized that normal and uncoupled EGG recordings have different
typical energy distribution throughout their wavelet transform coefficients.
Thus, the quality assessment parameters of compressed normal and uncoupled electrogastrograms
would be different,
since distinct amounts of energy would be used in the signal reconstruction.

The aim of this study is to present a method to quantitatively
detect mild and severe gastric electrical uncoupling using wavelet compression in a canine model.

\section{Methods}

\subsection{Experimental Setup}

After a laparotomy and the instalment of six pairs of internal
subserosal stainless steel wire electrodes into the antral gastric
wall of sixteen acute dogs (seven female and nine male), two complete circumferential cuts were
made fully dividing the organ into separate sections, but preserving the blood supply in each.
After each cut, the stomach was not anastomosed.
The sites of the cuts were selected to be distal to the
gastro-esophageal junction (first cut) and proximal to
gastro-esophageal junction (second cut). Consequently, by the
end of the experiment, these circular myotomies divided the
stomach in three parts of approximately equal dimensions.
Similar type of experimental work was described before~\cite{Mintchev:97,Mintchev:1998,Carre:2001}.

\begin{figure}
\centering
\begin{tabular}{cc}
\includegraphics[scale=0.85]{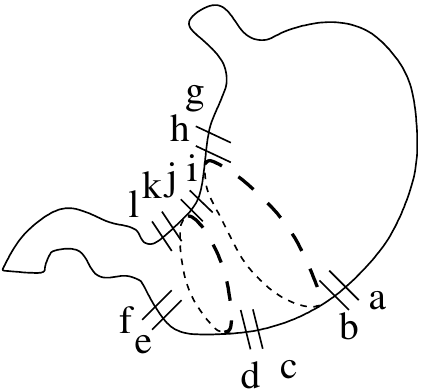}
&
\includegraphics[scale=0.85]{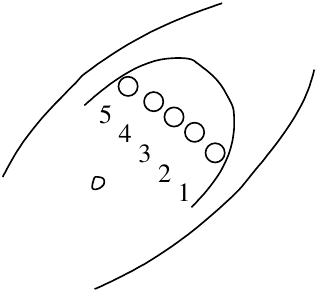}
\\
(a) & (b) \\
\input{electrode_table_gea.tex}
&
\input{electrode_table_egg.tex}
\\
(c) & (d)
\end{tabular}
\caption{Internal~(a) and cutaneous~(b) electrode positioning.
Various electrode combinations were used for the GEA~(c) and the EGG~(d) recordings.}
\label{egg:pos}
\end{figure}

After each incision, the abdominal wall was closed and five standard
disposable neonatal electrocardiography Ag-AgCl electrodes (Conmed,
Andover Medical, Haverhill, MA, USA) were collinearly placed on the
abdominal wall along the projection of the gastric axis. Additionally a
reference electrode was positioned in the area of the hip. Previous
studies demonstrated that this electrode configuration can be regarded as
optimal~\cite{Mirizzi:1983}. The five active electrodes were
grouped to provide eight bipolar EGG channels. Furthermore, six
stainless-steel wire electrodes implanted subserosally provided six
bipolar channels of internal GEA. However, in the present study only the
eight EGG recordings were processed, while the internal GEA channels were
used as a visual reference only to verify that normal electrical activity
was present. The electrode combination set and a diagram of the physical
location of the electrodes are depicted on Figure~\ref{egg:pos}.

Thirty-minute EGG recordings were performed in the three different states:
(i) basal state;
(ii) mild gastric electrical uncoupling (after the first cut);
and
(iii) severe gastric electrical uncoupling (after the second cut).

The captured EGG  signals were conditioned by a 0.02--0.2~\mbox{Hz}
low-pass
first order Butterworth active filter.
After amplification, 12-bit analog-to-digital conversion was performed
using a sampling frequency of 10~\mbox{Hz}
and
{\sc Labmaster 20009} 16-channel analog-to-digital converter
(Scientific Solutions, Vancouver, BC, Canada).
Thus, each half-hour recording generated 18{,}000 samples
per channel per state
(basal, mild uncoupling after the first cut, and severe uncoupling after the second cut)
per dog.

All experiments were approved by the Animal Welfare Committee and
the Ethics Committee at the Faculty of Medicine, University of Alberta
(Edmonton, Alberta, Canada).

\begin{figure}
\centering
\subfigure[]{\includegraphics[width=.45\linewidth]{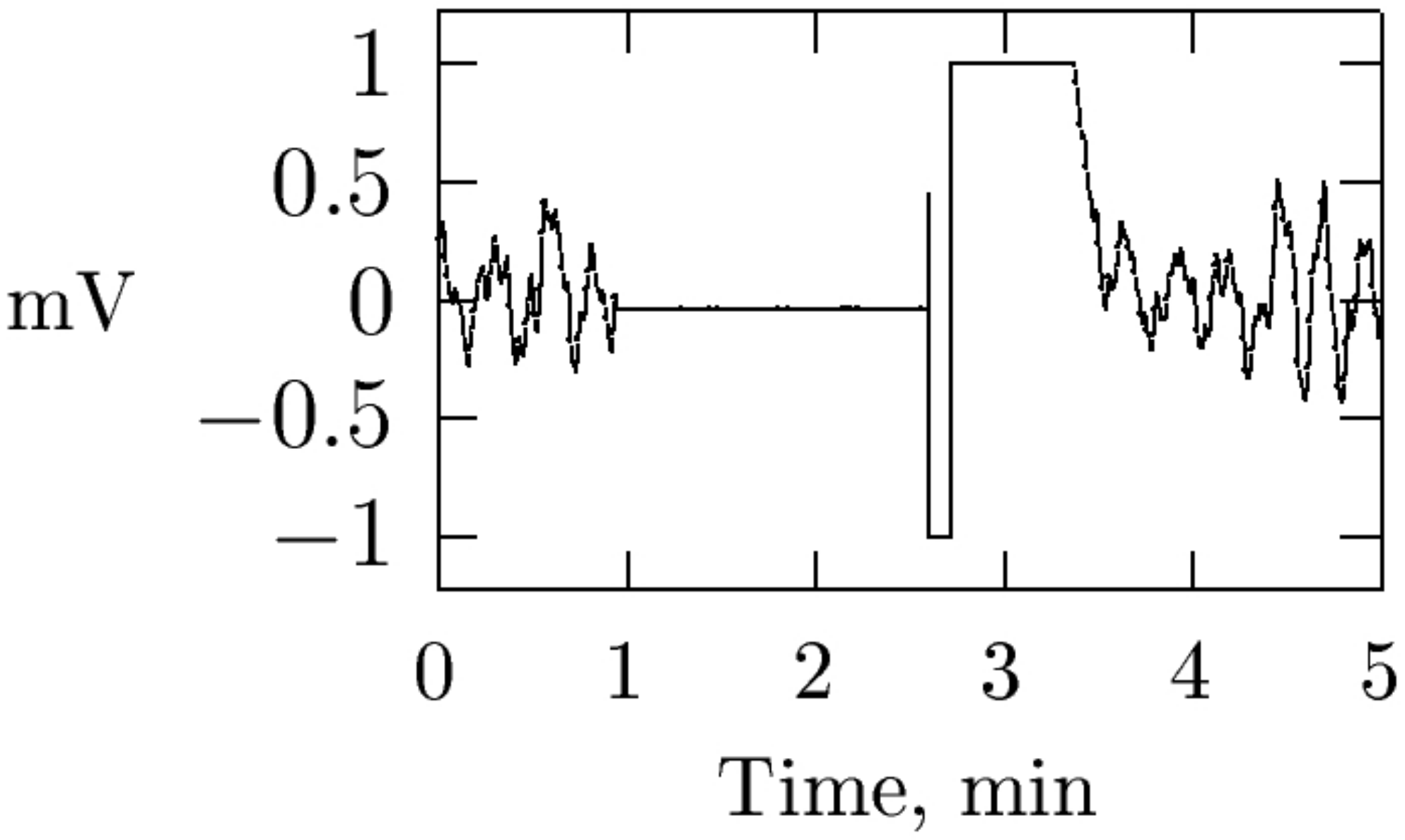}}
\subfigure[]{\includegraphics[width=.45\linewidth]{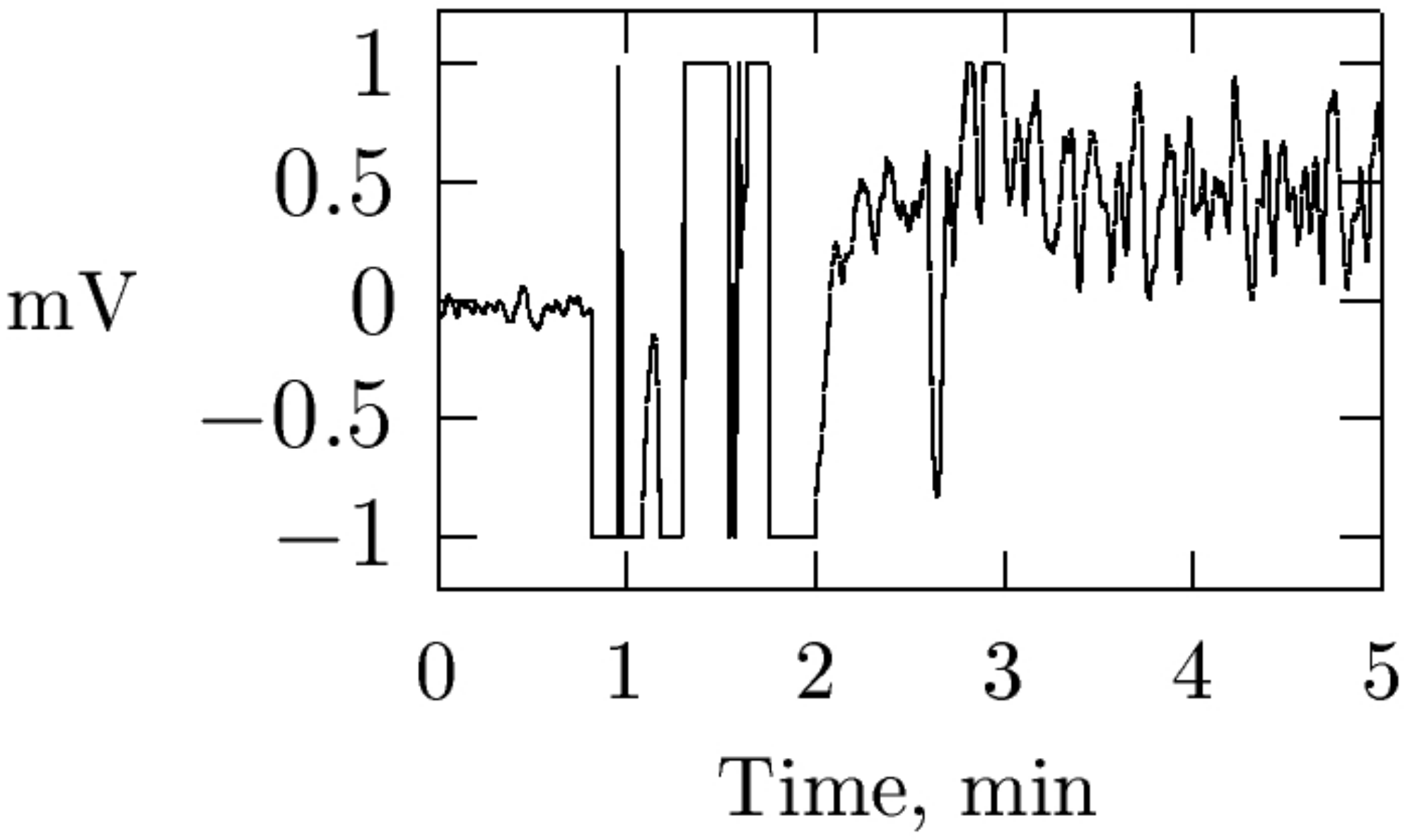}}\\
\subfigure[]{\includegraphics[width=.45\linewidth]{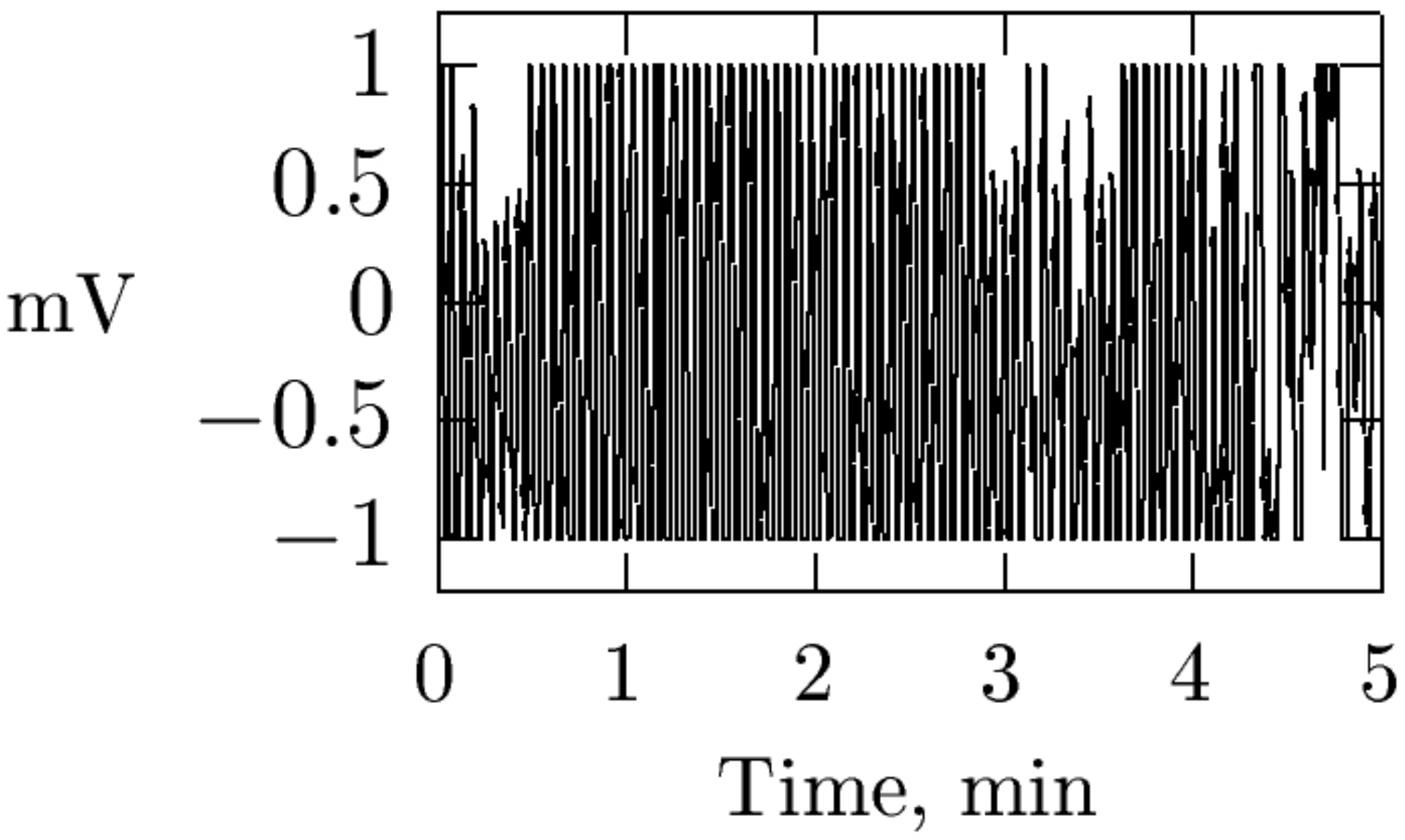}}
\subfigure[]{\includegraphics[width=.45\linewidth]{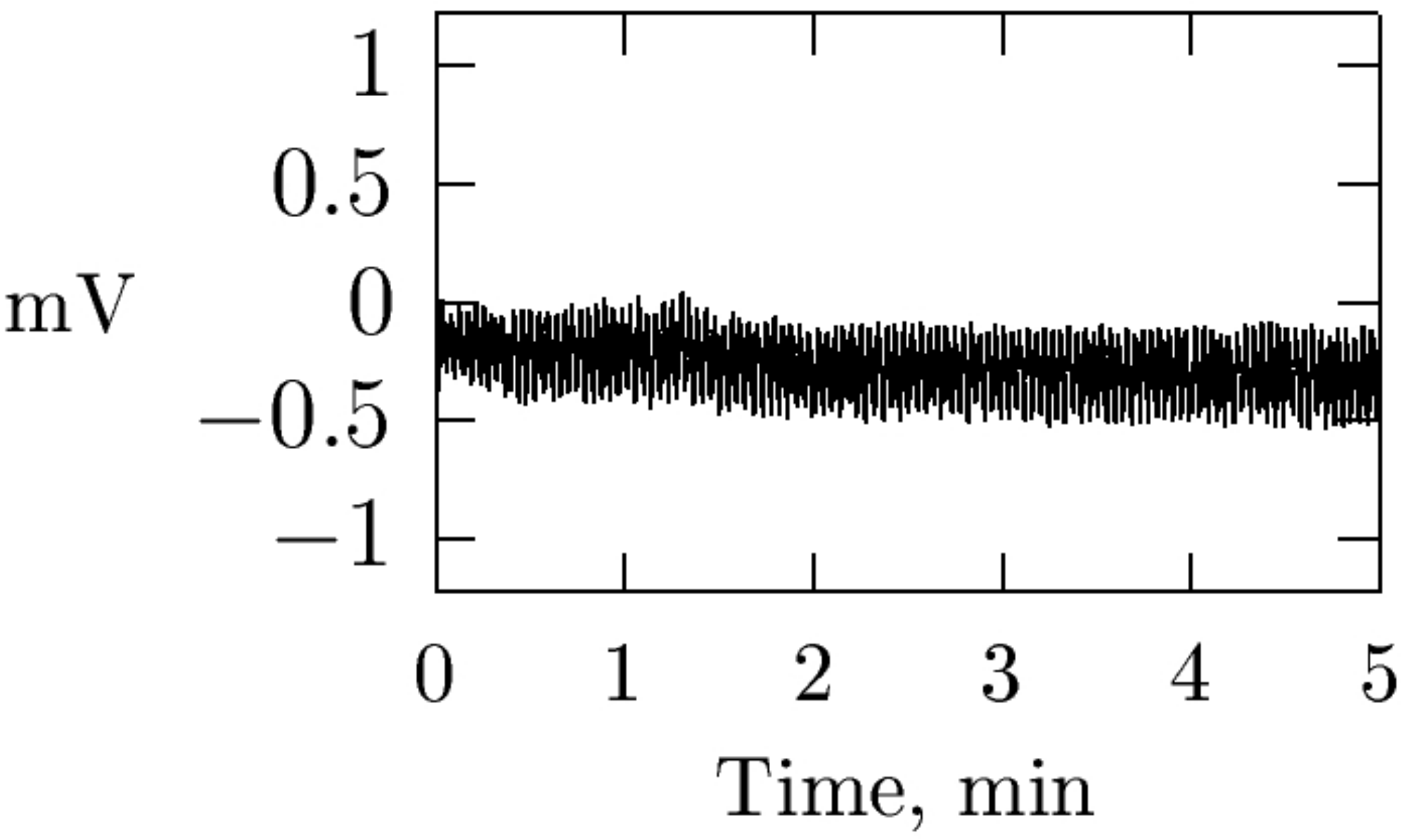}}
\caption{Examples of common artefacts found in raw EGG recordings:
loss of signal~(a),
offset change and saturation~(b),
noise and saturation~(c), and
lack of signal due to technological problems~(d).}
\label{artifacts}
\end{figure}

Since the recordings were of significant
duration,
the raw EGG data recordings
were intermittently contaminated with a multitude of artefacts,
including:
(i) motion artefacts;
(ii) spontaneous variations in electrode potentials;
(iii) respiration;
(iv) signal saturation  during recording;
(v) electrocardiac activity;
and
(vi) loss of signal during recording.
Usually these artefacts appeared simultaneously in all recording channels.
Figure~\ref{artifacts} depicts examples of various categories of
corrupted signals.
Some of these noisy patterns were visually evident (e.g., iv and~vi)
and could be easily identified and
discarded~\cite{Verhagen:1999}.
This practice has been recommended before in order to obtain a more reliable signal for subsequent
analysis~\cite{Parkman:03}.

\begin{figure}
\centering
\includegraphics[width=0.85\linewidth]{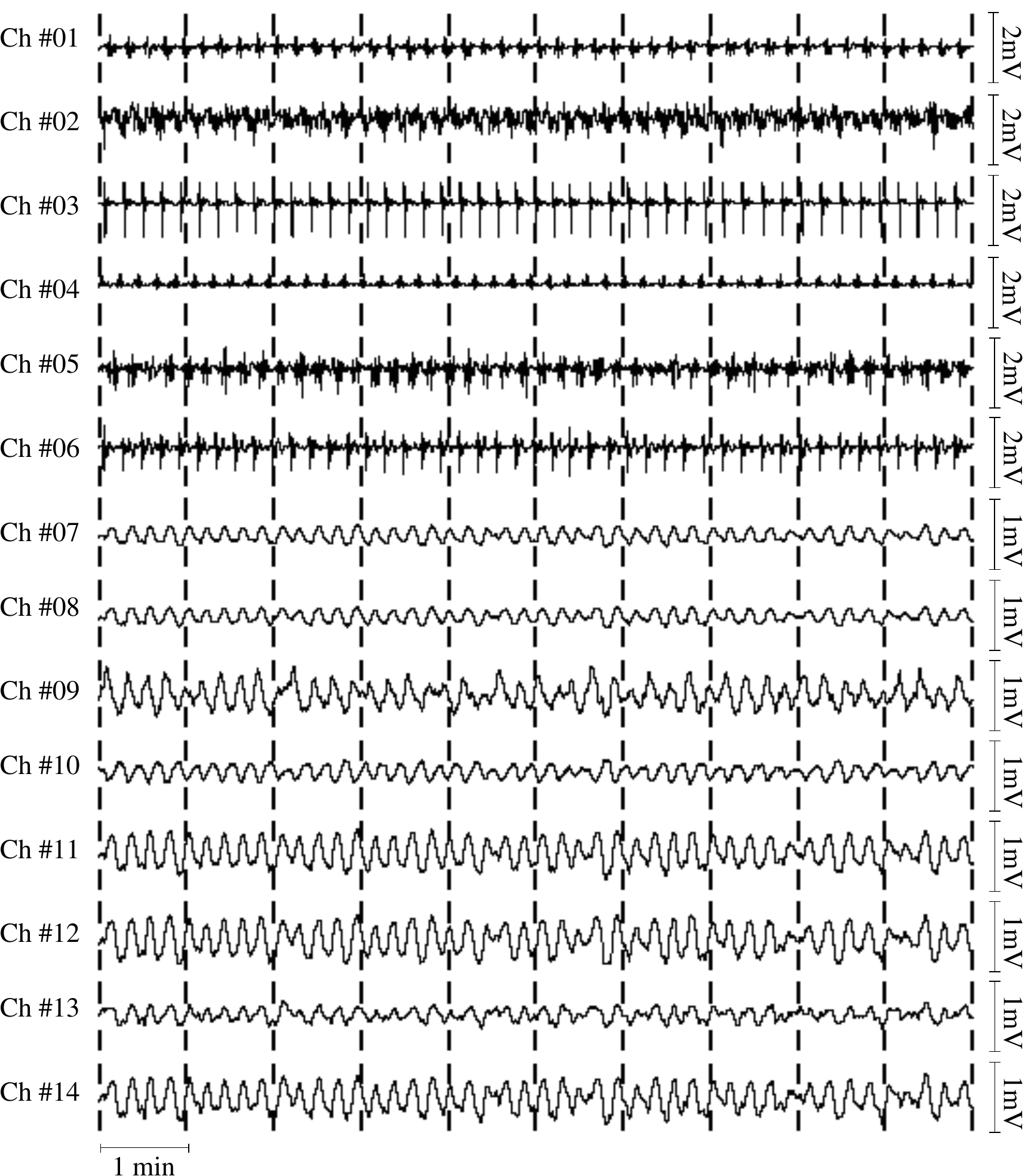}
\caption{Typical multichannel electrogastrographic tracings in the basal state. Channels 1--6 are internal GEA,
channels 7--14 are EGG.}
\label{normal_egg}
\end{figure}

Therefore, for each dog, a 10-minute time interval
of channel-synchronized data was manually selected.
These data were considered to be free from identifiable noise patterns.
Figure~\ref{normal_egg} shows a typical 10-minute  multichannel basal EGG recording.

\subsection{Signal Analysis}

\subsubsection{Wavelet Compression}

Wavelet theory suggests that it is possible to choose
a wavelet function $\psi(\cdot)$ that
generates an orthogonal basis in which a given signal
is
to be decomposed~\cite{Mallat:03}.
A continuous signal $x(t)$ has its wavelet transform coefficients
$c_{j,k}$
computed by
\begin{equation}
c_{j,k}
=
\int_{-\infty}^\infty x(t) \psi_{j,k}(t)\mathrm{d}t,
\end{equation}
where
the basis
$\psi_{j,k}(\cdot) = 2^{-j/2}\psi(2^{-j}\cdot - k)$
is controlled by a scale (dilation) integer index~$j$
and a
translation integer index~$k$.

Under some assumptions~\cite{Mallat:03}, these coefficients uniquely
represent $x(t)$, which can be reconstructed by the following wavelet series
\begin{equation}
\label{wavelet_reconstruction}
x(t) = \sum_{j=-\infty}^\infty\sum_{k=-\infty}^\infty c_{j,k} \psi_{j,k}(t).
\end{equation}
Equation~\ref{wavelet_reconstruction} represents
the synthesis equation or inverse wavelet transform~\cite{Unser2003}.

This transform has to be modified for
processing digitized signals.
Moreover, in
a discrete-time formalism
the direct definition of the wavelet transform
is computationally intensive.
Wavelet transforms are performed via the Fast Wavelet Transform using
Mallat's pyramid algorithm for decomposition (forward transform) and
reconstruction (inverse transform)~\cite{Mallat:03,Oliveira:04}.

\begin{figure}
\centering
\includegraphics[width=0.6\linewidth]{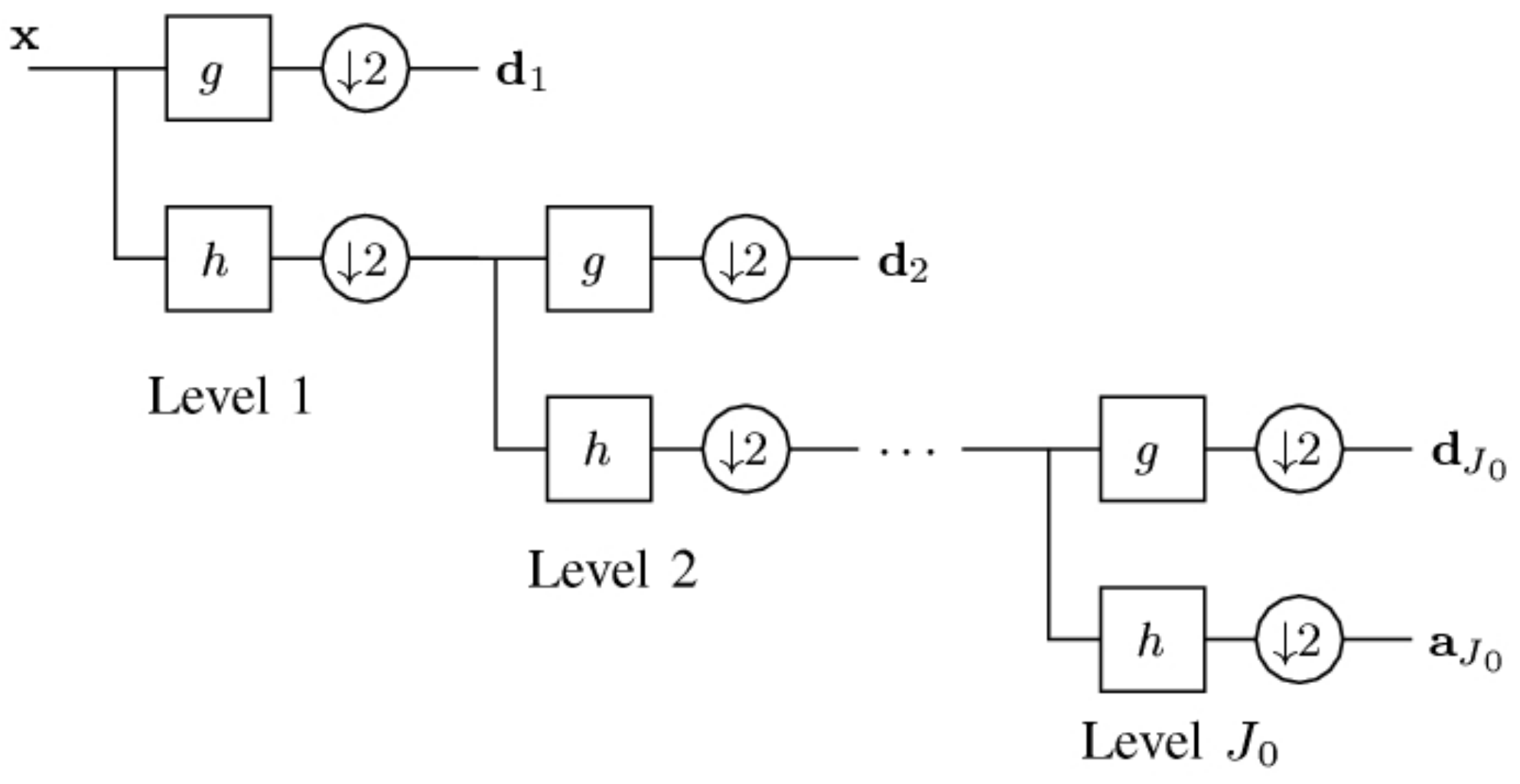}
\caption{Wavelet analysis filter bank. The signal is iteratively decomposed through
a filter bank to obtain its discrete wavelet transform.}
\label{filter_bank}
\end{figure}

Let $\mathbf{x}$ be a discrete signal with $N=2^J$ points
(a sampled version of the analog signal~$x(t)$).
The discrete wavelet transform (DWT) of $\mathbf{x}$ is computed in a recursive
cascade structure consisting of
decimators $\downarrow\!\!2$
and
complementing low-pass filter $h$ and high-pass filter $g$,
which are uniquely associated with a wavelet~\cite{Bratteli,MisiMisi00}.
Figure~\ref{filter_bank} depicts a diagram of the filter bank structure.

\begin{figure}
\centering
\subfigure[]{\includegraphics[width=0.85\linewidth]{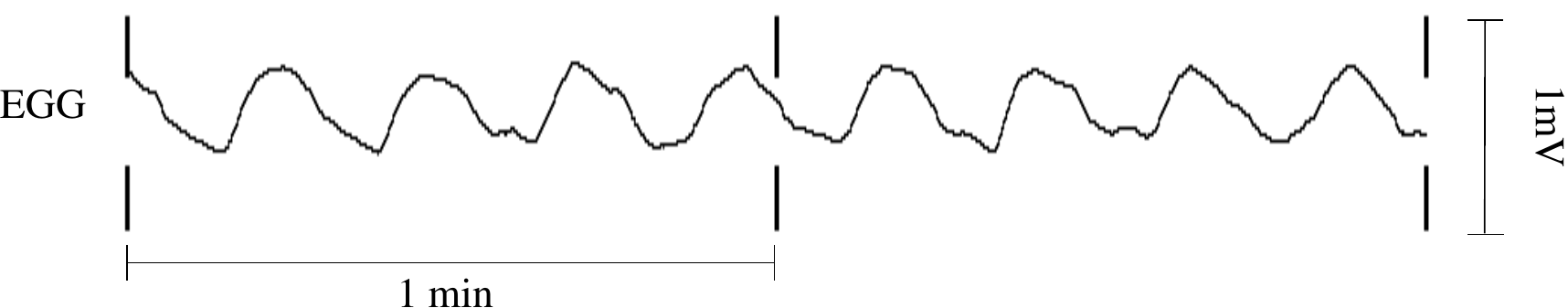}}\\
\subfigure[]{\includegraphics[width=0.85\linewidth]{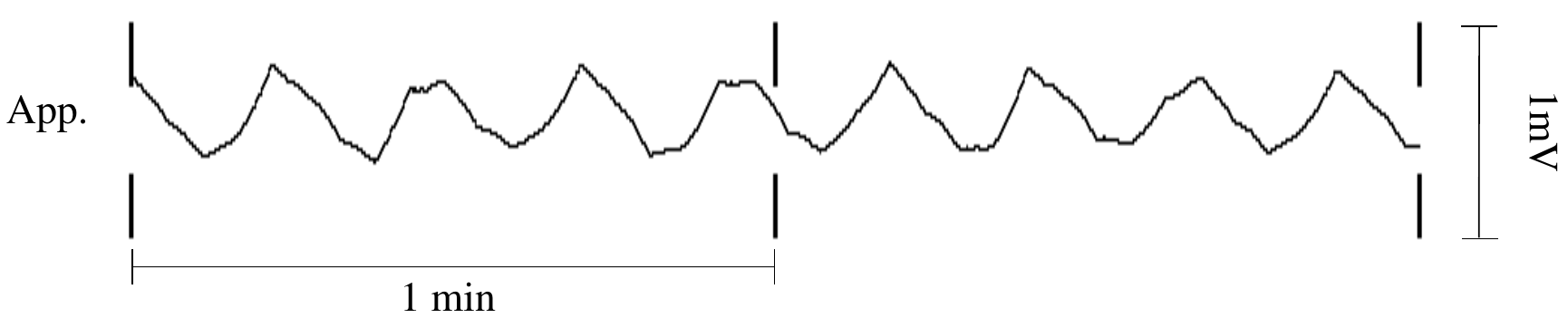}}\\
\subfigure[]{\includegraphics[width=0.85\linewidth]{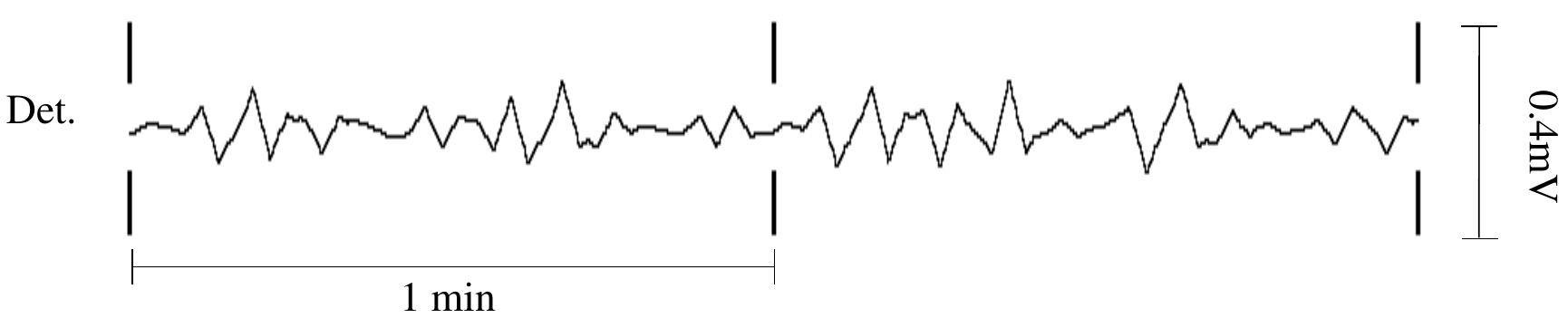}}\\
\caption{A 2-minute segment of a typical EGG signal in the basal state (a),
decomposed into approximation (b) and detail (c) signals after the sixth
iteration utilizing Mallat's pyramid algorithm. The wavelet utilized in this
analysis was Daubechies-2.}
\label{egg-app-det}
\end{figure}

At the end of the algorithm computation,
a set of vectors is obtained
\begin{equation}
\Big\{ \mathbf{d}_1, \mathbf{d}_2, \ldots, \mathbf{d}_j, \ldots, \mathbf{d}_{J_0}, \mathbf{a}_{J_0} \Big\},
\end{equation}
where~$J_0$ is
the number of decomposition scales of the DWT.
This set of approximation and detail vectors represents the DWT of the original signal.
Vectors~$\mathbf{d}_j$ contain the DWT detail coefficients of
the signal
in each scale~$j$.
As $j$ varies from~1 to $J_0$,
a finer or coarser
detail coefficient vector is obtained.
On the other hand, the vector~$\mathbf{a}_{J_0}$
contains the approximation coefficients of the signal at scale~$J_0$.
It should be noted that this recursive procedure can be iterated~$J$ times at most.
Usually, the procedure is iterated~$J_0< J$ times.
Depending on the choice of~$J_0$, a different set of
coefficients can be obtained.
Observe that the discrete signal~$\mathbf{x}$ and its DWT have the same length~$N$.
The inverse transformation can be performed using a similar recursive approach~\cite{MisiMisi00}.
Generally, a signal can be subject to various wavelet decompositions.
The analysis depends on
(i) the choice of wavelet (filters $h$ and $g$);
and
(ii) the number of decomposition levels (scales)~$J_0$.
Figure~\ref{egg-app-det} shows an EGG signal in the basal state along with the associated
approximation and detail signals, reconstructed from the respective
approximation and detail vectors obtained as a result of wavelet-based
decomposition utilizing Daubechies-2 wavelet~\cite{Mallat:03}.

A wavelet based compression scheme aims to satisfactorily represent
an original discrete signal~$\mathbf{x}$ with as few DWT coefficients
as possible~\cite{Vetterli:2001,Chagas:00,Lu:99,Watson:95}.
One simple and effective way of doing that is to discard
the coefficients that, under certain criteria, are considered
insignificant.
Consequently, the signal reconstruction
is based on a reduced set of coefficients~\cite{Vetterli:2001,Unser1996}.

In the present work the classic scheme for non-linear compression
was used~\cite{Vetterli:2001}.
This procedure considers an \emph{a posteriori} adaptive set,
which keeps  $M$
wavelet transform coefficients
that have the largest absolute values.
A hard thresholding was utilized to set the remaining coefficients to zero.
The number of coefficients $M$ to be retained was determined according
to the desired compression ratio $\CR$, which was defined by
\begin{equation}
\CR = \frac{N}{M},
\end{equation}
where $N$ and $M$ are the number of wavelet transform coefficients of
the original and the compressed signals, respectively.

Thus all three parameters
in a wavelet compression scheme have been defined:
(i) the number of scales;
(ii) the compression ratio;
and
(iii) the type of the wavelet.
However, before deriving
approaches to determine each of these parameters,
some additional issues need to be addressed.

\subsubsection{Measurement of Distortion}

The suggested assumption is
that the normal and uncoupled EGG recordings
have different typical energy distributions
throughout the wavelet coefficients.
Thus, depending on the signal (coupled or uncoupled), its energy
is distributed differently
among the wavelet transform coefficients.
If the energy of the signal is spread evenly
throughout the coefficients,
fixing the number of discarded coefficients
would result in greater distortions.
By contrast, if the energy of the signal is concentrated
in few wavelet coefficients,
the odds are that they would survive the compression
process, and
consequently, the reconstructed signal would be closer to the original.
Hence,
for a fixed compression ratio,
reconstructed signals
present different distortions,
depending on their energy distribution in
the wavelet domain.

Therefore, to further the present analysis, it is necessary to introduce
a quality assessment tool
to compare the original discrete signal~$\mathbf{x}$
with its reconstruction~$\tilde{\mathbf{x}}$.
Several measures that allow the
evaluation of the effect of compression schemes
have been suggested~\cite{Besar:00}.
However, one of the most commonly used is
the Percent Root-mean-square Difference
($\prd$)~\cite{Chagas:00,Besar:00,Lu:99},
which was utilized in the present study
as a measure of distortion
in the compression scheme.
The $\prd$ of two signals, $\mathbf{x}$ and $\tilde{\mathbf{x}}$,
both of length~$N$,
is defined by:
\begin{equation}
\prd(\mathbf{x},\tilde{\mathbf{x}})
=
\sqrt{\frac{\sum_{i=0}^{N-1}  (x_i - \tilde{x}_i)^2 }{\sum_{i=0}^{N-1} x_i^2}}
\times
100 \%.
\end{equation}

\subsection{Statistical Analysis}
\label{section_stats}

After computing the quality assessment parameter ($\prd$) for each canine EGG signal,
the signals were grouped according
to the state of the dog (basal, mild uncoupling, and severe uncoupling) and the recording channel.
Significant statistical difference ($p<0.05$)
between the
$\prd$ values
of compressed EGG signals
obtained from
the basal state and mild or severe gastric electrical uncoupling
for a given channel was sought.

Since the EGG signals of the three groups were acquired from the
same sixteen dogs, the assumption of data independence
could be questioned.
Consequently, a small-sample inference using
straightforward $t$-statistics might not be appropriate~\cite{Snedecor97}.

Paired statistical approaches, such as the Paired Difference test~\cite{Snedecor97},
were utilized to
compare the $\prd$ between two given groups of signals.
Therefore, paired differences of the
$\prd$'s ($\Delta\prd$) were computed.
Since Paired Difference test can only be used when
the relative frequency distribution of the
population of differences is normal,
the nonparametric Wilcoxon signed rank test
could also be taken in consideration as an alternative~\cite{Snedecor97}.

To check the assumption of normal distribution of the $\Delta\prd$ samples,
the Lilliefors (Kolmogorov-Smirnov) test of normality was utilized~\cite{Gonzalez1977}.
Whenever Lilliefors test verified that the
difference between the samples satisfied a normal distribution,
the Paired Difference test was utilized to compare the two groups.
Otherwise, the less restrictive Wilcoxon signed rank test was employed.

\subsection{Choice of Parameters}

\subsubsection{Number of Scales}
\label{number_of_scales}

In order to select the number of scales $J_0 \in \{1, \ldots, J\}$
of the wavelet transform decomposition,
the following criterion was introduced:~$J_0$
was chosen so that the coarsest approximation
scale
had a pseudo-frequency close to the
canine EGG dominant frequency $f_c$
of 4--6 cycles per minute~\cite{Mintchev:2000}.

The pseudo-frequency $f_{\mbox{pseudo}}$ of a given scale~$j$ is
\begin{equation}
f_{\mbox{pseudo}}
=
\frac{f_\psi}{j \cdot T_s},\qquad j=1,\ldots,J,
\end{equation}
where~$T_s$ is the sampling period (0.1~\mbox{s}) and~$f_\psi$
is the center frequency of a wavelet
(the frequency that maximizes the magnitude of the Fourier transform of the wavelet)~\cite{MisiMisi00}.
Consequently,
the scale~$J_0$ was selected which minimized the difference
$(f_{\mbox{pseudo}} - f_c)$.
Table~\ref{pseudo} shows the number of decomposition levels
for some common wavelets.

\begin{table}
\centering
\caption{\label{pseudo}Number of decomposition scales $J_0$ for some wavelets.}

\begin{tabular}{@{}*{2}{c}}
\br
Wavelet      & $J_0$ \cr
\mr
Daubechies-2 & 6     \cr
Daubechies-3 & 7     \cr
Coiflet-1    & 7     \cr
\br
\end{tabular}
\end{table}

\subsubsection{Compression Ratio}
\label{cr_section}

The selection of the compression ratio
was performed by
choosing the ratio that maximized the number of channels
which
exhibited statistically significant difference,
when comparing the basal to the uncoupled groups of signals.
In order to facilitate the selection process,
a plot was constructed relating
the compression ratio
to
the percentage of channels in which statistically significant
difference was observed.
This calculation was performed for two situations:
(i) comparing basal state to mild electrical uncoupling;
and
(ii) comparing basal state to severe electrical uncoupling.
Thus,
the values of the compression ratio
that delivered maximal percentage of detecting channels in both comparison cases
were selected.

In this study, three well-known wavelets were utilized,
namely, Daubechies-2, -3, and Coiflet-1.
The number of scales utilized was set accordingly (see Table~\ref{pseudo}).

\subsubsection{Choice of Wavelet}

Although there are numerous issues concerning
the choice of wavelet for signal analysis~\cite{Chapa2000},
generally,
a wavelet is better suited to a class of signals
if the latter can be represented by
as few wavelet coefficients as possible~\cite{Mallat:03,Vetterli:2001}.
Thus, wavelets which resemble the waveshape of the signal under analysis are often selected.
In the context of the present study,
a wavelet was sought that minimized
the $\prd$ between the original EGG signal and its reconstruction
for a fixed compression ratio.
If,
 for a given wavelet,
   the $\prd$ associated with a compressed signal was minimal,
then
 the surviving coefficients were considered to be
 better representing the original signal.
Therefore,
the selected wavelet was also more effectively ``matched''
to the signal under analysis when compared to other wavelets in consideration.
However, the abundance of wavelets~\cite{Bratteli}
makes such approach prohibitive.
As a result, some constraints on the choice of  wavelet were introduced.

It is well known that wavelets can be generated from
discrete finite impulse response (FIR) filters~\cite{MisiMisi00}.
In the present work, the analysis was limited to
wavelets generated by FIR
filters with length no greater than six coefficients.
In this subset of wavelets one may find Haar, Daubechies-2, Daubechies-3, and Coiflet-1 wavelets,
to name the most popular ones~\cite{Mallat:03}.

\begin{figure}
\centering
\includegraphics[width=0.35\linewidth]{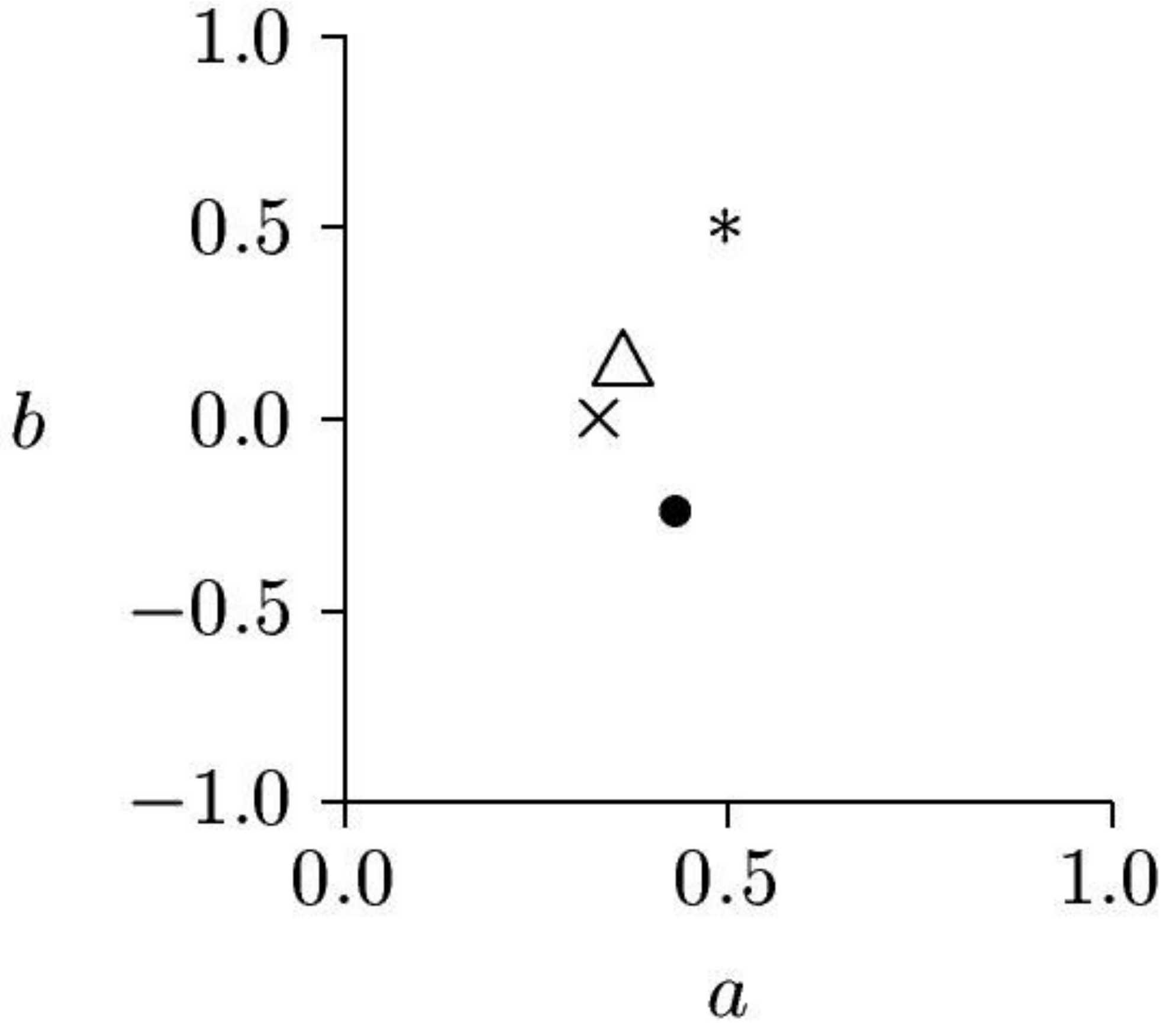}
\caption{Parameterization plane (the axes are normalized by $\pi$).
The coordinate points
that correspond to Haar ($\ast$),
Daubechies-2 ($\times$),
Daubechies-3 ($\bullet$),
and
Coiflet-1 ($\triangle$) wavelets are shown.}
\label{ab_plane}
\end{figure}

This restriction is quite convenient, since all FIR filters of length up to six
that can be utilized to generate wavelets have simple parameterizations
of their coefficients~\cite{Tew1992,Zhou1993}.
For example, Pollen parameterization of 6-tap wavelet filters~\cite{Tew1992}
has two independent variables $(a,b)\in[-\pi,\pi]\times [-\pi,\pi]$.
Varying these two parameters,
a filter that generates a new wavelet can be defined.
Consequently, the Pollen parameterization defines a plane on which every
point is connected to a wavelet~\cite{Bratteli}.
In Figure~\ref{ab_plane}, the parameterization plane is partially
depicted and some wavelet positions are denoted.

Using the discussed compression scheme,
one can compute a~$\prd$ value of a signal for each wavelet
generated from a point with coordinates~$(a,b)$ on the
parameterization plane.
Doing so, a surface can be defined by the points~$(a,b,\prd)$.
Thus, the minima of this surface correspond to the
point coordinates~$(a,b)$ that generate a wavelet with good ``matching''
properties, since the $\prd$ values at these minima are small.
On the other hand, the maxima of this surface indicate higher values of
$\prd$, consequently the reconstructed signal from the compression
is a poorer representation of the original.

\subsubsection{Testing the Methodology}

\begin{figure}
\centering
\includegraphics[width=0.5\linewidth]{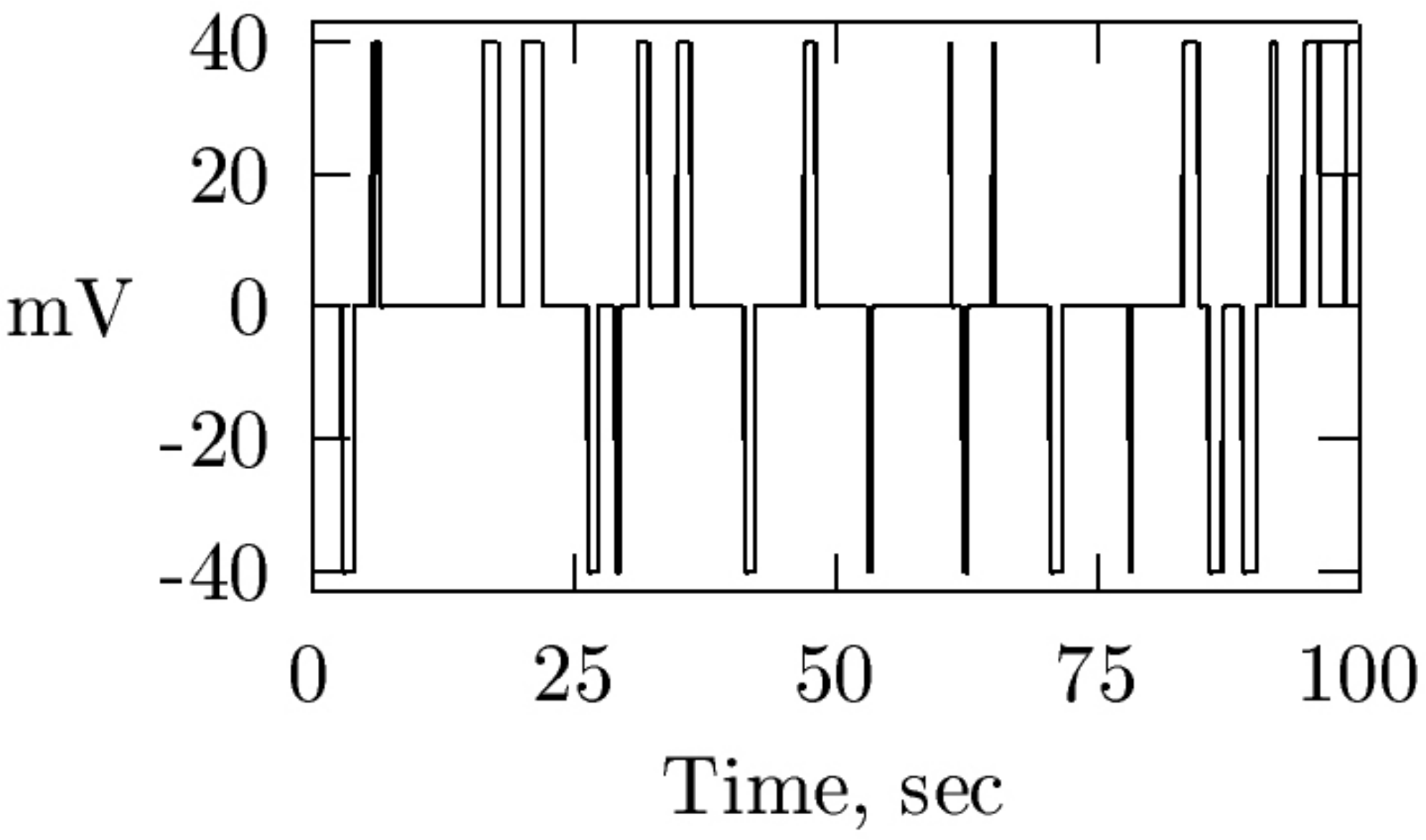}
\caption{Square-wave test signal of random polarity.}
\label{staircase}
\end{figure}

\begin{figure}
\centering
\includegraphics[width=.5\linewidth]{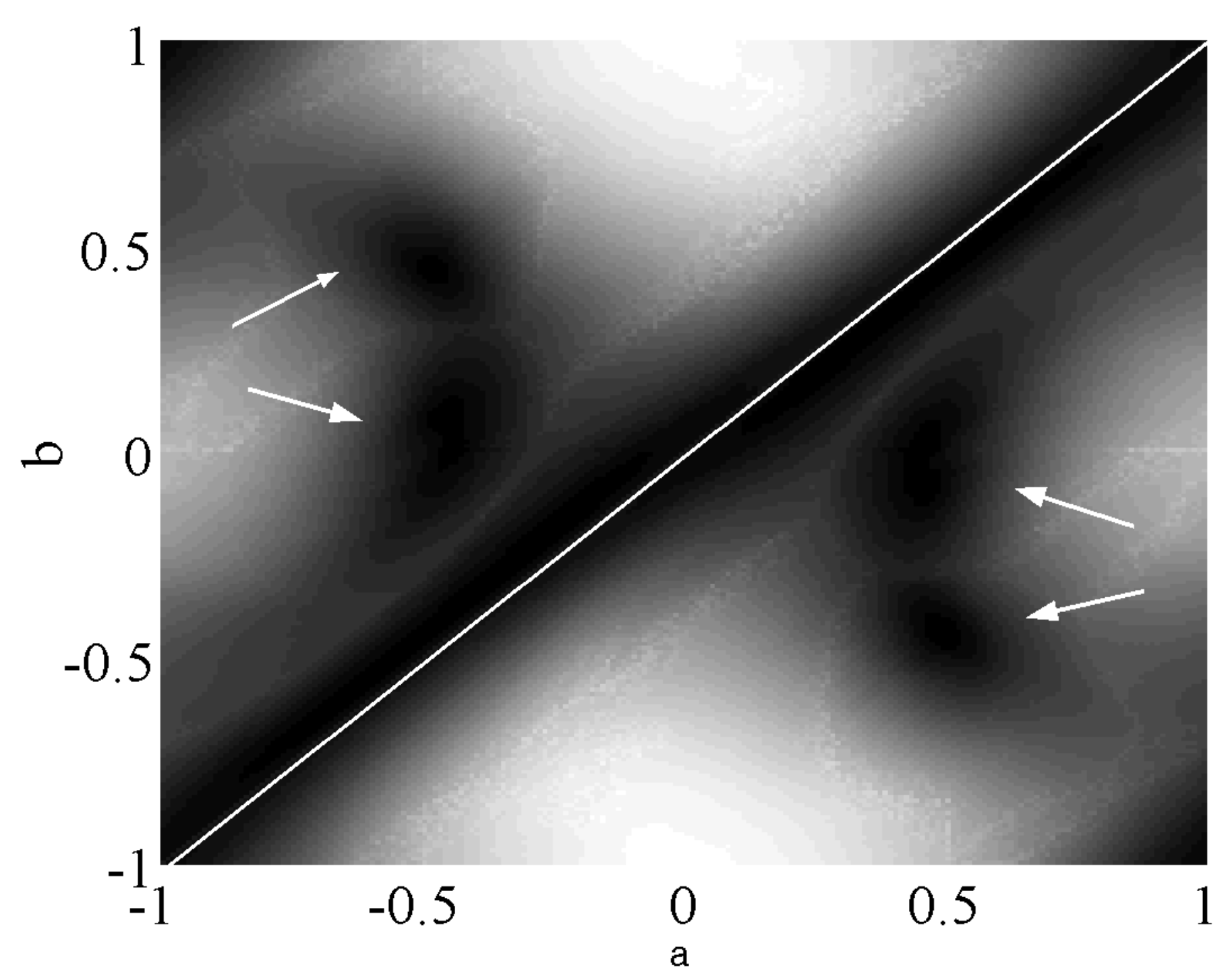}
\caption{$\prd$ surface on the parameterization plane
resulting from processing the square-wave test signal
with compression ration of~3.
A gray scale is used to represent the $\prd$ values.
The darker areas are the minima of the $\prd$ surface
and coincide with the regions that generate Haar wavelets:
the marked diagonal, $(0.5,-0.5)$,  $(0.5,0)$, $(-0.5,0.5)$, and $(-0.5,0)$ (indicated by the arrows).
The axes are normalized by~$\pi$.}
\label{stair_prd}
\end{figure}

\begin{figure}
\centering
\subfigure[]{\includegraphics[width=.45\linewidth]{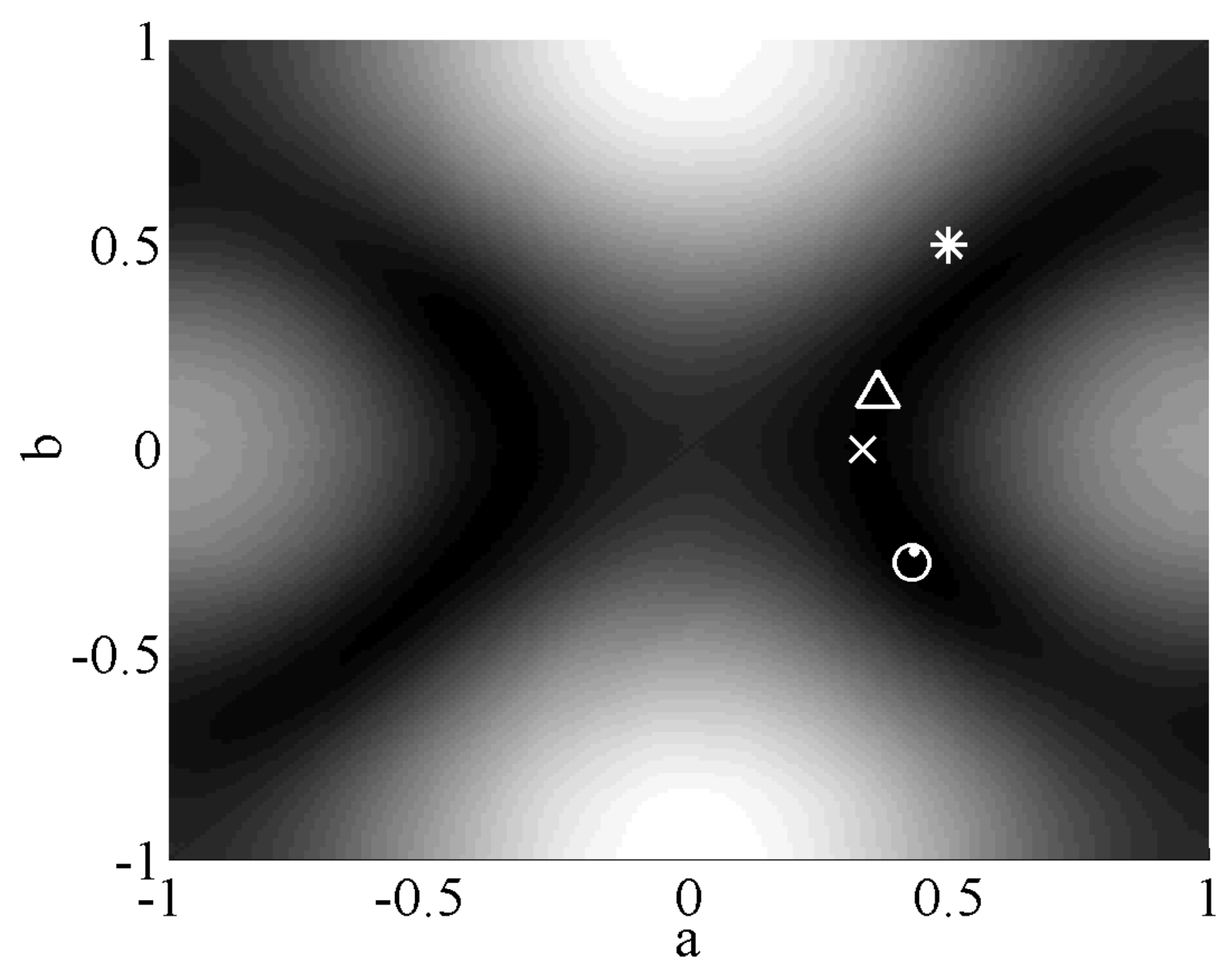}}
\subfigure[]{\includegraphics[width=.45\linewidth]{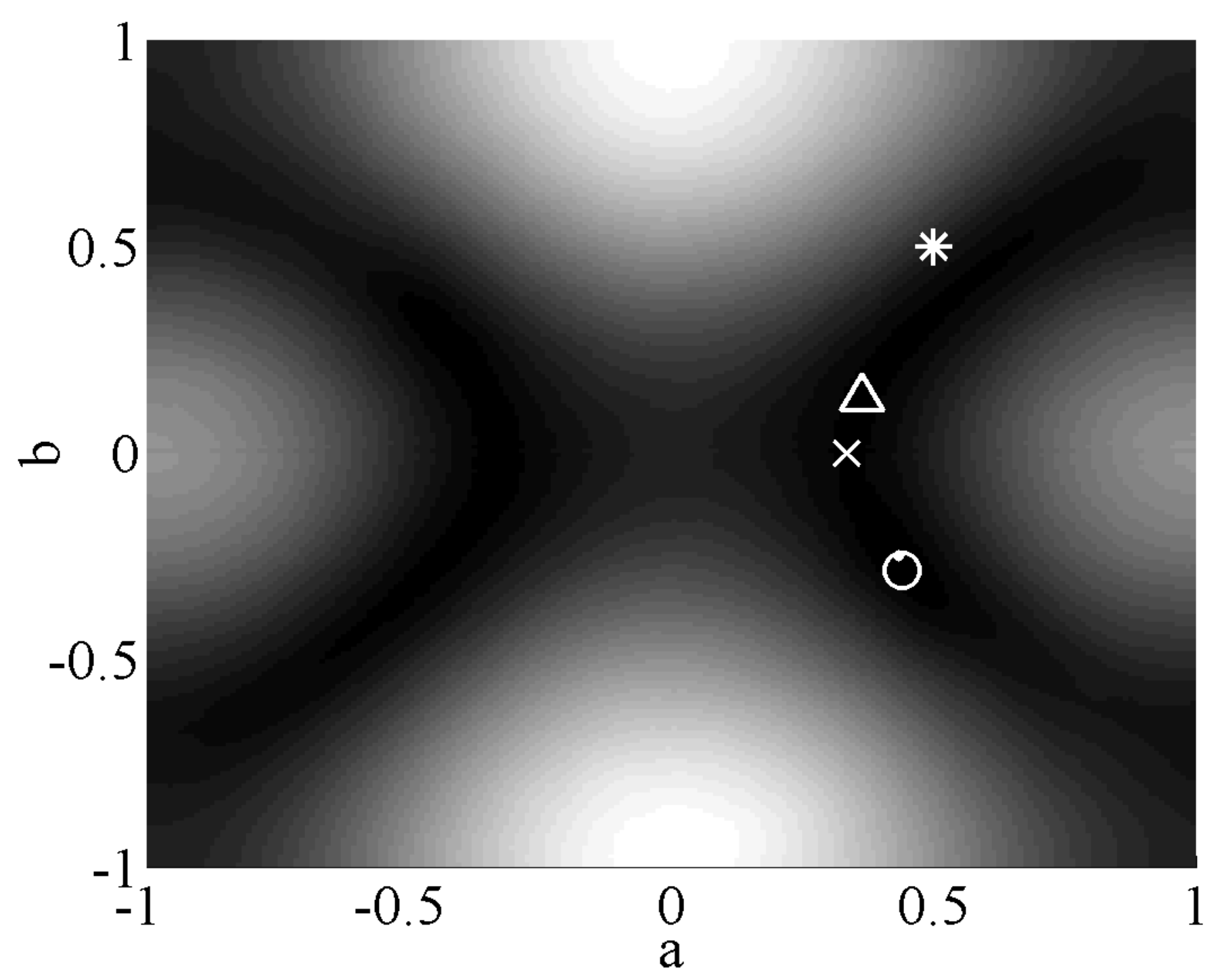}}
\caption{Plot generated after computing the $\prd$
surface for all possible wavelets on the parameterization plane
for two canine EGG signals in the basal state.
The minimum value is depicted by a circle ($\circ$).
The coordinate points
that correspond to Haar ($\ast$),
Daubechies-2 ($\times$),
Daubechies-3 ($\bullet$),
and
Coiflet-1 ($\triangle$) wavelets are shown.
The axes are normalized by~$\pi$.}
\label{prd-plane}
\end{figure}

To verify the validity of the proposed methodology, it was applied
to a square wave signal shown in Figure~\ref{staircase}. It has
been previously shown that the Haar wavelet matches this kind of
signal and can offer a good representation~\cite{Oliveira:04}.
Setting the compression ratio to~3 and the number of scales to~6,
a $\prd$ surface on the parameterization plane was constructed,
resulting in the plot shown in Figure~\ref{stair_prd}. A minimum
of this surface occurs at the point  $(\pi/2, \pi/2)$, which
corresponds exactly to the Haar wavelet. Other minima are located
on $(\pi/2, -\pi/2)$, $(\pi/2, 0)$, $(-\pi/2, \pi/2)$, $(-\pi/2,
0)$, and on the diagonal. All these regions also generate
Haar-like wavelets. Therefore, the proposed procedure of finding
the minima of the $\prd$ surface could be applied to every EGG
recording in the basal state, and a wavelet that would ``match''
normal EGG recordings could be defined. As a result, a set of
point coordinates $(a_i, b_i)$ could be determined on the
parameterization plane, which minimizes the $\prd$ for each normal
canine EGG recording~$i$. Figure~\ref{prd-plane} shows typical
surfaces generated for basal canine EGG signals~\cite{Gopi1994}.

Taking the mean value of the minima,
the best wavelet parameterization $(a^\ast, b^\ast)$ could be defined.
Thus,
$(a^\ast,b^\ast)$
generates a wavelet that on average
``matches'' best the normal EGG recordings.

\section{Results}

\subsection{Determination of Parameters}

\subsubsection{Compression Ratio}
\begin{figure}
\centering
\subfigure[Basal vs. Mild Electrical Uncoupling]{\includegraphics[width=0.41\linewidth]{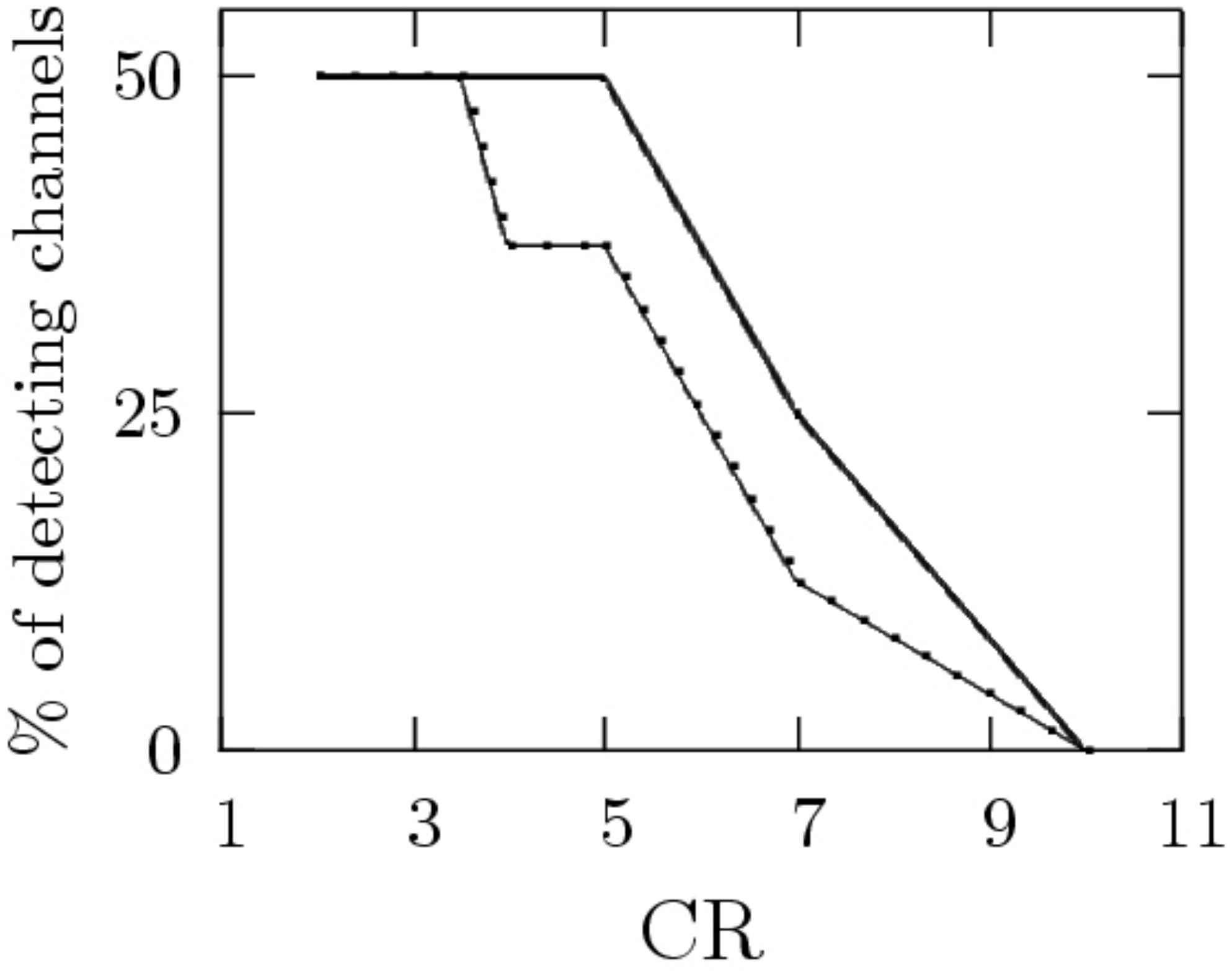}}
\quad
\subfigure[Basal vs. Severe Electrical Uncoupling]{\includegraphics[width=0.41\linewidth]{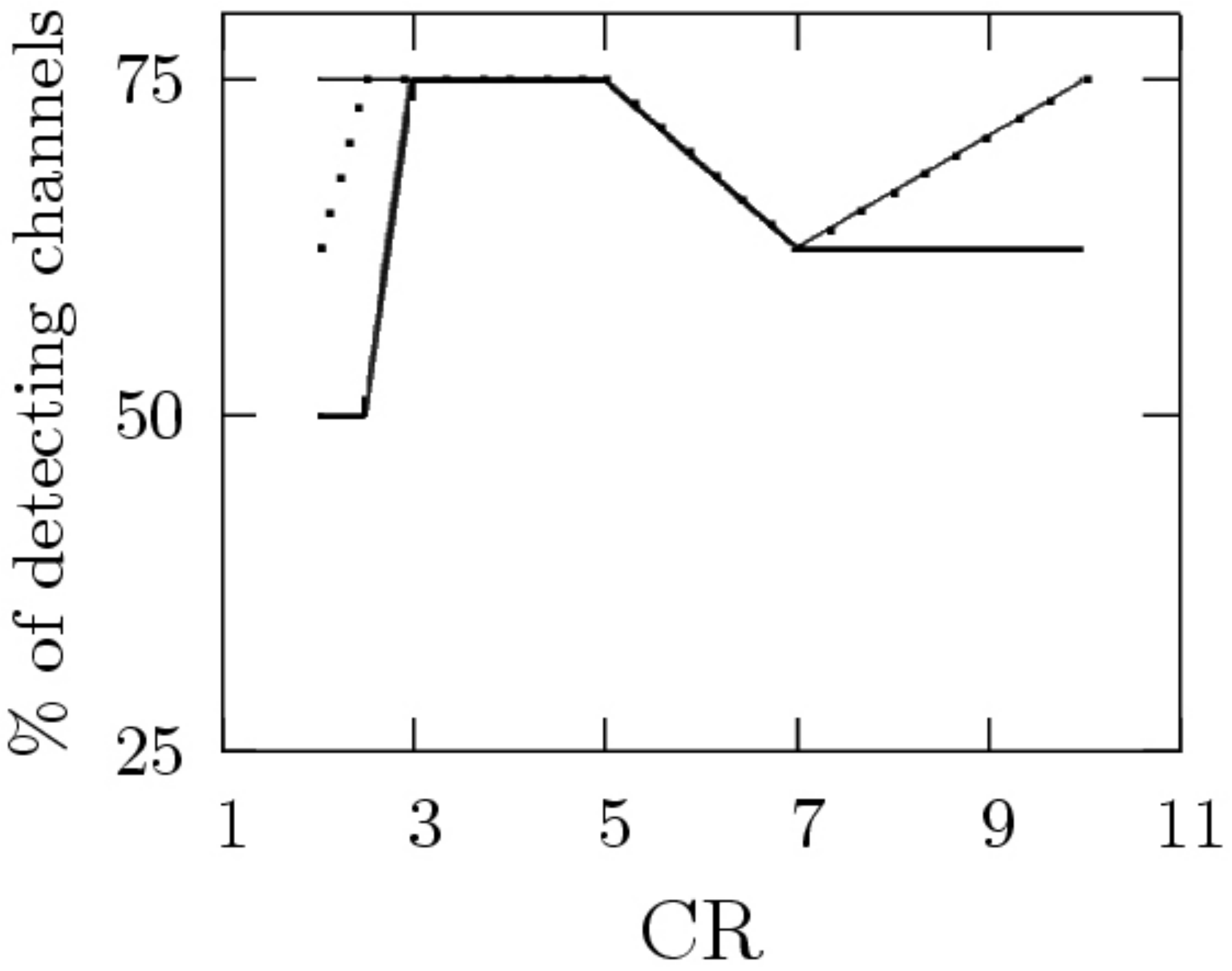}}
\caption{Compression ratio vs percentage of
channels in which successful detection occurred ($p<0.05$).
Each curve displays the analysis for a specific wavelet:
Daubechies-2 (thin solid line),
Daubechies-3 (solid line),
and
Coiflet-1 (dotted line).}
\label{cr_vs_sens}
\end{figure}

The $\prd$ values for all EGG recordings were computed
using some selected wavelets (Daubechies-2, Daubechies-3, and Coiflet-1)
and several compression ratios.
Haar wavelet was not taken into consideration because it is not continuous,
and offers poor approximations for smooth functions like the~EGG signals~\cite{Mallat:03}.
Figure~\ref{cr_vs_sens} shows the results of this procedure.

It is interesting to observe that percentage of channels
in which statistically significant difference was found
is highly dependent on the choice of compression ratio
when comparing basal state and mild gastric electrical uncoupling groups.
As the compression ratio increased,
the resulting $\prd$ sets from
the two groups
became statistically less distinct,
deteriorating the ability to discriminate between them.
Figure~\ref{cr_vs_sens}(a) illustrates this behaviour.

On the other hand, the proposed algorithm is more prone to detect
severe electrical uncoupling.
Significant statistical difference ($p<0.044$)
between
basal state and severe electrical uncoupling
was observed in up to~6 out of~8 channels.
Moreover, the percentage of channels where
significance
was consistently observed was relatively
constant with respect to
the change of compression ratio (Figure~\ref{cr_vs_sens}(b)).

Evidently, depending on the choice of wavelet,
compression ratio range between 3 and 5
offered the best outcomes for both cases.
Therefore, compression ratio of~3 was selected,
which was the smallest compression ratio
with which both plots achieved maximum values.

\subsubsection{Wavelet}

\begin{figure}
\centering
\includegraphics[width=0.5\linewidth]{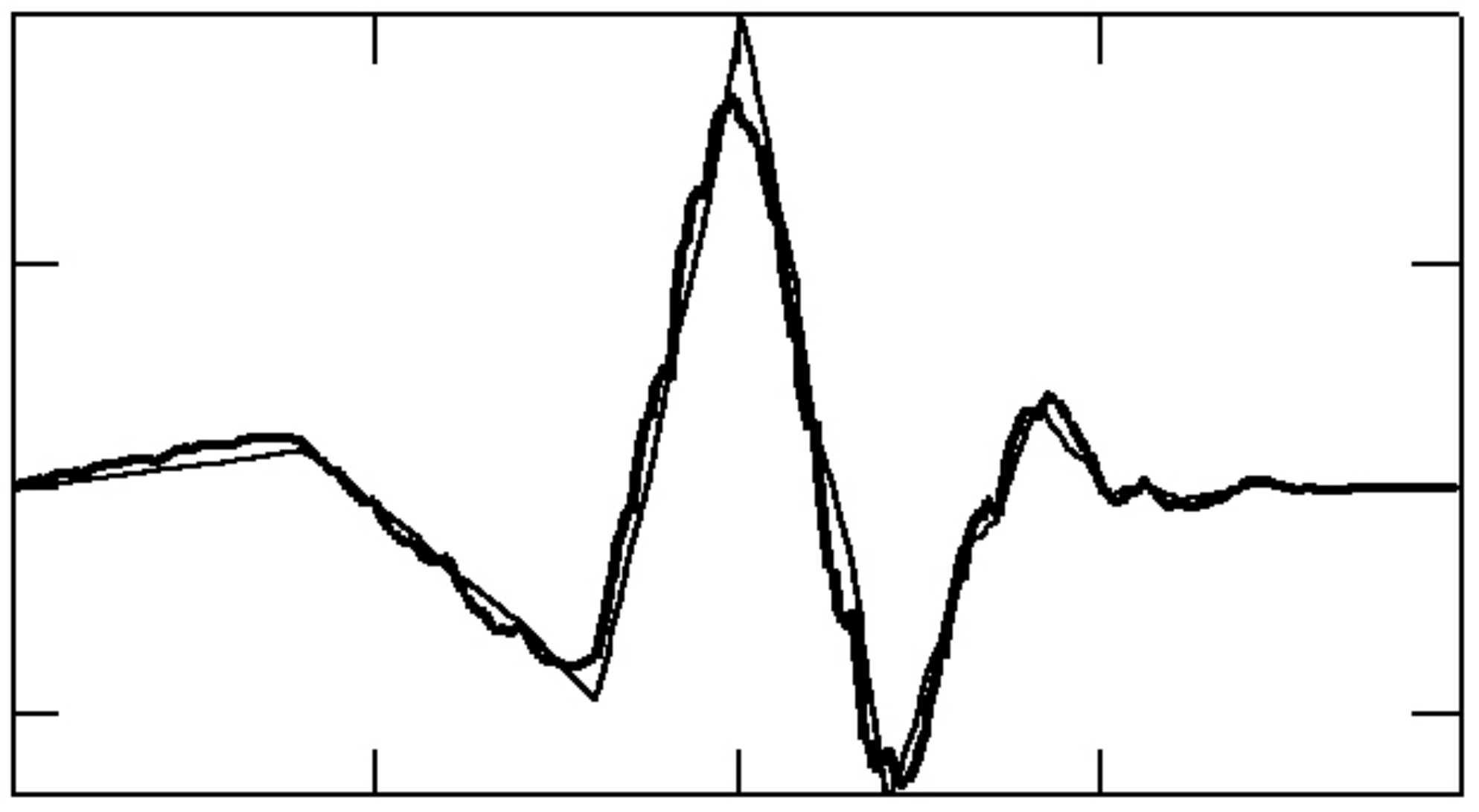}
\caption{Comparison between the optimal wavelet (solid bold curve) obtained using the
derived parameters~$a^\ast$ and~$b^\ast$, and
the standard Daubechies-3 wavelet (solid thin curve).
The similarity between the two ones is clearly evident.}
\label{best_wavelet}
\end{figure}

Setting the compression ratio to three, the specific values of $a^\ast = 0.43$
and $b^\ast = -0.26$ were determined, which were associated
with
the wavelet depicted in Figure~\ref{best_wavelet} (bold curve).
It is worth mentioning that the proposed wavelet was
very similar to the classic Daubechies-3 wavelet.
Since Daubechies-3 is
(i)~very similar to the proposed wavelet;
and
(ii)~easily available in many software packages
(e.g., {\sc Matlab} (The Mathworks, Inc., Natick, MA, USA)),
it was chosen instead of the proposed wavelet.
Previous empirical findings~\cite{Ryu:02} confirm this observation.

\subsection{Statistics}

Tables~\ref{ab_db3_cr3}, \ref{ac_db3_cr3} and~\ref{bc_db3_cr3}
summarize the results when
Daubechies-3 wavelet and compression ratio of 3 were utilized.

\begin{table}
\centering
\caption{\label{ab_db3_cr3}Comparison between the $\prd$ values
in the basal state and mild gastric electrical uncoupling.}

\begin{tabular}{@{}*{6}{c}}
\br
Channel&Statistics&$\Delta\prd$ Mean&$\Delta\prd$ SD& Significant?&$p$-value\cr
\mr
7 & Student   & 0.982161 & 1.844312 & No  & 0.058226 \cr
8 & Student   & 0.566675 & 0.679417 & Yes & 0.010919 \cr
9 & Student   & 0.728762 & 1.137274 & Yes & 0.026378 \cr
10 & Student  & 0.192864 & 1.314660 & No  & 0.592370 \cr
11 & Wilcoxon & 1.347208 & 1.741442 & Yes & 0.000854 \cr
12 & Student  & 1.186993 & 1.948825 & Yes & 0.033388 \cr
13 & Wilcoxon & 0.648649 & 1.532822 & No  & 0.187622 \cr
14 & Student  & 0.424109 & 1.643240 & No  & 0.334474 \cr
\br
\end{tabular}
\end{table}

\begin{table}
\centering
\caption{\label{ac_db3_cr3}Comparison between the $\prd$ values in the
basal state and severe gastric electrical uncoupling.}

\begin{tabular}{@{}*{6}{c}}
\br
Channel&Statistics&$\Delta\prd$ Mean&$\Delta\prd$ SD& Significant?&$p$-value\cr
\mr
7 & Student  & 0.924724 & 1.416927 & Yes & 0.029661 \cr
8 & Student  & 0.755751 & 0.763789 & Yes & 0.005647 \cr
9 & Student  & 0.596236 & 1.029897 & Yes & 0.049474 \cr
10 & Student & 0.306322 & 1.505244 & No  & 0.428371 \cr
11 & Student & 0.813283 & 1.044165 & Yes & 0.015799 \cr
12 & Student & 0.972415 & 0.899240 & Yes & 0.001386 \cr
13 & Student & 0.634854 & 1.066521 & Yes & 0.044225 \cr
14 & Student & 0.467240 & 0.983946 & No  & 0.112567 \cr
\br
\end{tabular}
\end{table}

\begin{table}
\centering
\caption{\label{bc_db3_cr3}Comparison between the $\prd$ values in
mild and severe gastric electrical uncoupling state.}

\begin{tabular}{@{}*{6}{c}}
\br
Channel&Statistics&$\Delta\prd$ Mean&$\Delta\prd$ SD& Significant?&$p$-value\cr
\mr
7 & Student  & \phantom{-} 0.011820  & 1.365449 & No & 0.973727 \cr
8 & Student  & \phantom{-} 0.074321  & 0.803788 & No & 0.744598 \cr
9 & Student  & $-$0.191332  & 1.317089 & No & 0.582588 \cr
10 & Student & $-$0.169222  & 1.481669 & No & 0.676127 \cr
11 & Student & $-$0.441275  & 1.605023 & No & 0.322378 \cr
12 & Student & $-$0.304088  & 1.756426 & No & 0.513433 \cr
13 & Student & \phantom{-} 0.044421  & 1.093672 & No & 0.877249 \cr
14 & Student & $-$0.208983  & 1.242392 & No & 0.540001 \cr
\br
\end{tabular}
\end{table}

\section{Discussion}

In the employed model of gastric electrical uncoupling,
as the myotomies were performed
the electrical power produced by the intrinsic gastric generator
was divided into two or three generators of lower electric power,
according to the number of the circumferential cuts.
Consequently, the
signal-to-noise ratio of the recorded signal decreased
and the influence of various external sources of disturbance became more expressive.
Therefore, the energy of the EGG signal became
more distributed throughout the wavelet coefficients, when compared to the signal
in the basal state (single gastric generator).
Due to the design of the proposed method,
as the energy was distributed in more coefficients,
the reconstruction error increased.

The evaluation of the reconstruction error of  compressed canine EGG signals
using the proposed wavelet technique was able to discriminate
between induced
severe gastric electrical uncoupling
and the basal state
in 75\%~of the EGG channels.
The detection capability of the procedure was reduced when
comparing the control group to the mild gastric uncoupling group
but nevertheless,
significant differences were observed in 4~out of 8~EGG channels
(see Tables~\ref{ab_db3_cr3} and~\ref{ac_db3_cr3}).
This increase in sensitivity when detecting severe uncoupling is expected
and reinforce previous findings~\cite{Mintchev:97,SanMiguel:1999,Carre:2001}.
Channels 8, 11 and 12 were recognized as the ones that presented
the best overall performance in detecting uncoupling signals
($p$-values ranging from $0.0009$ to $0.01$).
The site of such channels oversees the
mid to distal corpus of the
epigastric region.
These electrode locations probably encompass
fully electrical activity from all uncoupling regions.
This observation emphasizes the appropriateness of using
a multichannel EGG recording system.

A careful examination of Table~\ref{ac_db3_cr3} reveals that channels~9 and~13
have only borderline significance.
A more conservative interpretation of the data (e.g., $p<0.01$)
would reduce the ability of the method to detect mild uncoupling
to a single channel~(\#11),
which,
however,
still represents a significant improvement in recognizing mild uncoupling,
compared to previous findings~\cite{SanMiguel:1999}.

Mild and severe gastric electrical uncoupling groups were statistically indistinguishable
($p>0.05$).
This indicates that, according to the proposed technique,
EGG signals contain insufficient information
to assess the level of gastric electrical uncoupling.
In addition, when comparing mild and severe uncoupling,
a non-consistent tendency of change in the reconstruction error
was observed.
This can be noted from the fluctuation of the sign of
the mean reconstruction error in Table~\ref{bc_db3_cr3}.

Previous works~\cite{Carre:2001,SanMiguel:1999,Mintchev:1998,Mintchev:97}
suggested that determinism, randomness, and
chaos are dynamically blended in the EGG signals. The present
investigation indicates that a single signal processing approach might be
insufficient to provide an accurate assessment method for EGG recordings.
Because of the complex dynamic nature of the EGG signal, various methods
such as classical time-frequency analysis, wavelet theory, non-linear
techniques, and biomagnetic field pattern recognition could be combined to
provide sufficent data for an informed diagnostic decision. Indeed, such
combined approach could be a key to converting electrogastrography into a
reliable clinical tool.

\section{Conclusion}

A new wavelet analysis approach based on signal compression
is proposed for the detection of mild and severe gastric electrical
uncoupling in a canine model.
Combining the results of the suggested wavelet compression scheme
with a multichannel recording system and
a comprehensive electrode configuration mapping the
abdominal wall along the projection of the gastric axis,
it was demonstrated that significant
difference between the basal state and gastric electrical uncoupling
can be detected in particular EGG channels.
These channels are
usually located along the abdominal projection of the
mid to distal corpus axis.

\section*{Acknowledgments}

This work was supported by
the Natural Sciences and Engineering Research Council of Canada (NSERC)
and the National Council for Scientific and Technological Development (CNPq, Brazil).

{\small
\bibliographystyle{IEEEtran}
\bibliography{wavelet}
}

\end{document}

%% file: electrode_table_gea.tex
{ \footnotesize
\begin{tabular}{cc}
\br
Channel &  \parbox{2cm}{\centering Electrode Combination} \\
\mr
1      & a--b \\
2      & c--d \\
3      & e--f \\
4      & g--h \\
5      & i--j \\
6      & k--l \\
\br
\end{tabular}}

%% file: electrode_table_egg.tex
{ \footnotesize
\begin{tabular}{cc}
\br
Channel &  \parbox{2cm}{\centering Electrode Combination} \\
\mr
7       & 1--2 \\
8       & 2--3 \\
9       & 3--4 \\
10      & 4--5 \\
11      & 1--3 \\
12      & 1--4 \\
13      & 2--5 \\
14      & 1--5 \\
\br
\end{tabular}}

%% file: waiacmogeu-cleaned.bbl
\begin{thebibliography}{10}
\providecommand{\url}[1]{#1}
\csname url@samestyle\endcsname
\providecommand{\newblock}{\relax}
\providecommand{\bibinfo}[2]{#2}
\providecommand{\BIBentrySTDinterwordspacing}{\spaceskip=0pt\relax}
\providecommand{\BIBentryALTinterwordstretchfactor}{4}
\providecommand{\BIBentryALTinterwordspacing}{\spaceskip=\fontdimen2\font plus
\BIBentryALTinterwordstretchfactor\fontdimen3\font minus
  \fontdimen4\font\relax}
\providecommand{\BIBforeignlanguage}[2]{{%
\expandafter\ifx\csname l@#1\endcsname\relax
\typeout{** WARNING: IEEEtran.bst: No hyphenation pattern has been}%
\typeout{** loaded for the language `#1'. Using the pattern for}%
\typeout{** the default language instead.}%
\else
\language=\csname l@#1\endcsname
\fi
#2}}
\providecommand{\BIBdecl}{\relax}
\BIBdecl

\bibitem{Szurszewski1981}
J.~H. Szurszewski, \emph{Physiology of the Gastrointestinal Tract}.\hskip 1em
  plus 0.5em minus 0.4em\relax New York: Raven Press, 1981, ch. Electrical
  Basis for Gastrointestinal Motility, pp. 1435--1466.

\bibitem{Daniel94}
E.~E. Daniel, B.~L. Bardakjian, J.~D. Huizinga, and N.~E. Diamant, ``Relaxation
  oscillator and core conductor models are needed for understanding of {GI}
  electrical activities,'' \emph{American Journal of Physiology:
  Gastrointestinal and Liver Physiology}, vol. 226, no.~3, pp. 475--482, 1994.

\bibitem{Borto98}
M.~Bortolotti, ``Electrogastrography: A seductive promise, only partially
  kept,'' \emph{The American Journal of Gastroenterology}, vol.~93, no.~10, pp.
  1791--1794, Oct. 1998.

\bibitem{Chen95}
J.~D.~Z. Chen, J.~Pan, and R.~W. McCallum, ``Clinical significance of gastric
  myoelectrical dysrhythmias,'' \emph{Digestive Diseases and Sciences},
  vol.~13, no.~5, pp. 275--290, Sep. 1995.

\bibitem{Chen93}
J.~D.~Z. Chen, W.~R. {Stewart Jr.}, and R.~W. McCallum, ``Spectral analysis of
  episodic rhythmic variations in cutaneous electrogastrogram,'' \emph{IEEE
  Transactions on Biomedical Engineering}, vol.~40, no.~2, pp. 128--135, Feb.
  1993.

\bibitem{Parkman:03}
H.~P. Parkman, W.~L. Hasler, J.~L. Barnett, and E.~Y. Eaker,
  ``Electrogastrography: a document prepared by the gastric section of the
  american motility society clinical {GI} motility testing task force,''
  \emph{Neurogastroenterology and Motility}, vol.~15, pp. 89--102, 2003.

\bibitem{YouChey1984}
C.~H. You and W.~Y. Chey, ``Study of electromechanical activity of the stomach
  in humans and dogs with particular attention to tachygastria,''
  \emph{Gastroenterology}, vol.~86, pp. 1460--1468, 1986.

\bibitem{Mintchev97}
M.~P. Mintchev and K.~L. Bowes, ``Do increased electrogastrographic frequencies
  always correspond to internal tachygastria?'' \emph{Annals of Biomedical
  Engineering}, vol.~25, pp. 1052--1058, 1997.

\bibitem{Bradshaw03}
L.~A. Bradshaw, A.~G. Myers, A.~Redmond, J.~P. Wikswo, and W.~O. Richards,
  ``Biomagnetic detection of gastric electrical activity in normal and
  vagotomized rabbits,'' \emph{Neurogastroenterology and Motility}, vol.~15,
  no.~5, pp. 475--482, Oct. 2003.

\bibitem{Publi89}
N.~G. Publicover and K.~M. Sanders, ``Are relaxation oscillators an appropriate
  model of gastrointestinal electrical activity?'' \emph{American Journal of
  Physiology: Gastrointestinal and Liver Physiology}, vol. 256, no.~2, pp.
  265--274, 1989.

\bibitem{Smout80}
A.~J. P.~M. Smout, E.~J. {Van Der Schee}, and J.~L. Grashuis, ``What is
  measured in electrogastrography?'' \emph{Digestive Diseases and Sciences},
  vol.~25, no.~3, pp. 179--186, Mar. 1980.

\bibitem{Mintchev:97}
M.~P. Mintchev, S.~J. Otto, and K.~L. Bowes, ``Electrogastrography can
  recognize gastric electrical uncoupling in dogs,'' \emph{Gastroenterology},
  vol. 112, pp. 2006--2011, 1997.

\bibitem{Mintchev:1998}
M.~P. Mintchev, A.~Stickel, and K.~L. Bowes, ``Dynamics of the level of
  randomness in gastric electrical activity,'' \emph{Digestive Diseases and
  Sciences}, vol.~43, no.~5, pp. 953--956, May 1998.

\bibitem{Carre:2001}
J.-Y. Carr\'e, A.~{H{\o}st-Madsen}, K.~L. Bowes, and M.~P. Mintchev, ``Dynamics
  of the level of deterministic chaos associated with gastric electrical
  uncoupling in dogs,'' \emph{Medical \& Biological Engineering \& Computing},
  vol.~39, pp. 322--329, 2001.

\bibitem{Bradshaw99}
L.~A. Bradshaw, J.~K. Ladipo, J.~P. {Wikswo, Jr.}, and W.~O. Richards, ``The
  human vector magnetogastrogram and magnetoenterogram,'' \emph{IEEE
  Transactions on Biomedical Engineering}, vol.~46, no.~8, pp. 959--970, Aug.
  1999.

\bibitem{Bradshaw2003}
L.~A. Bradshaw, A.~Myers, J.~P. Wikswo, and W.~O. Richards, ``A spatio-temporal
  dipole simulation of gastrointestinal magnetic fields,'' \emph{IEEE
  Transactions on Biomedical Engineering}, vol.~50, no.~7, pp. 836--847, Jul.
  2003.

\bibitem{Xie98}
X.~Xie and H.~H. Sun, ``Sinusoidal time-frequency wavelet family and its
  application in electrogastrographic signal analysis,'' in \emph{Proceedings
  of the 20th Annual International Conference of the IEEE Engineering in
  Medicine and Biology Society}, vol.~3, Hong Kong, China, 1998, pp.
  1450--1453.

\bibitem{Ryu:02}
C.~Ryu, K.~Nam, S.~Kim, and D.~Kim, ``Comparison of digital filters with
  wavelet multiresolution filter for electrogastrogram,'' in \emph{Proceedings
  of the Second Joint BMES/EMBS Conference}, Houston, TX, Oct. 2000, pp.
  137--138.

\bibitem{Liang96}
J.~Liang, J.~C. Cheung, and J.~D.~Z. Chen, ``Noise detection and denoising on
  electrogastrography using nonorthogonal multiresolution wavelet analysis,''
  in \emph{Proceedings of the 18th Annual International Conference of the IEEE
  Engineering in Medicine and Biology Society}, vol.~3, Amsterdam, Netherlands,
  1996, pp. 1039--1040.

\bibitem{Liang02}
H.~Liang and Z.~Lin, ``Stimulus artifact cancellation in the serosal recordings
  of gastric myoelectric activity using wavelet transform,'' \emph{IEEE
  Transactions on Biomedical Engineering}, vol.~49, no.~7, pp. 681--688, Jul.
  2002.

\bibitem{Qiao96}
W.~Qiao, H.~H. Sun, W.~Y. Chey, and K.~Y. Lee, ``Continuous wavelet analysis as
  an aid in the representation and interpretation of electrogastrographic
  signals,'' in \emph{Proceedings of the Fifteenth Southern Biomedical
  Engineering Conference}, Dayton, USA, Mar. 1996, pp. 140--141.

\bibitem{Mirizzi:1983}
N.~Mirizzi and U.~Scafoglieri, ``Optimal direction of the electrogastrographic
  signal in man,'' \emph{Medical \& Biological Engineering \& Computing},
  vol.~21, no.~4, pp. 385--389, Jul. 1983.

\bibitem{Verhagen:1999}
M.~A. M.~T. Verhagen, L.~J. {Van Schelven}, M.~Samsom, and A.~J. P.~M. Smout,
  ``Pitfalls in the analysis of electrogastrographic recordings,''
  \emph{Gastroenterology}, vol. 117, no.~2, pp. 453--460, 1999.

\bibitem{Mallat:03}
S.~G. Mallat, \emph{A Wavelet Tour of Signal Processing}, 2nd~ed.\hskip 1em
  plus 0.5em minus 0.4em\relax Academic Press, 1999.

\bibitem{Unser2003}
M.~Unser and T.~Blu, ``Wavelet theory demystified,'' \emph{IEEE Transactions on
  Signal Processing}, vol.~51, no.~2, pp. 470--483, Feb. 2003.

\bibitem{Oliveira:04}
H.~M. {de Oliveira}, \emph{An\'alise de Sinais para Engenheiros: Uma Abordagem
  Via Wavelets}.\hskip 1em plus 0.5em minus 0.4em\relax S\~ao Paulo: Manole
  Pub., 2004.

\bibitem{Bratteli}
O.~Bratteli and P.~E.~T. Jorgensen, \emph{Wavelets Through A Looking Glass: The
  World of the Spectrum}.\hskip 1em plus 0.5em minus 0.4em\relax Boston:
  Birkh\"auser, 2002.

\bibitem{MisiMisi00}
M.~Misiti, Y.~Misiti, G.~Oppenheim, and J.-M. Poggi, \emph{Wavelet Toolbox
  User's Guide}, 2nd~ed.\hskip 1em plus 0.5em minus 0.4em\relax New York: The
  MathWorks, Inc., 2000.

\bibitem{Vetterli:2001}
M.~Vetterli, ``Wavelets, approximation, and compression,'' \emph{IEEE Signal
  Processing Magazine}, vol.~5, pp. 59--73, Sep. 2001.

\bibitem{Chagas:00}
A.~Chagas, E.~{Da Silva}, and J.~Nadal, ``{ECG} data compression using
  wavelets,'' in \emph{Computers in Cardiology}, Sep. 2000, pp. 423--426.

\bibitem{Lu:99}
Z.~Lu, D.~Y. Kim, and W.~Pearlman, ``{ECG} signal compression with a new
  wavelet method,'' in \emph{Proceedings of the First Joint BMES/EMBS
  Conference}, Atlanta, GA, Oct. 1999, p. 955.

\bibitem{Watson:95}
M.~J. Watson, A.~Liakopoulos, D.~Brzakovic, and C.~Georgakis, ``Wavelet
  techniques in the compression of process data,'' in \emph{Proceedings of the
  American Control Conference}, Seattle, USA, Jun. 1995, pp. 1265--1269.

\bibitem{Unser1996}
M.~Unser and A.~Aldroubi, ``A review of wavelets in biomedical applications,''
  \emph{Proceedings of the IEEE}, vol.~84, no.~4, pp. 626--638, Apr. 1996.

\bibitem{Besar:00}
R.~Besar, C.~Eswaran, S.~Sahib, and R.~J. Simpson, ``On the choice of the
  wavelets for {ECG} data compression,'' in \emph{Proceedings of the
  International Conference on Acoustics, Speech, and Signal Processing},
  Istanbul, Turkey, Jun. 2000, pp. 1011--1014.

\bibitem{Snedecor97}
G.~W. Snedecor and W.~G. Cochran, \emph{Statistical Methods}, 6th~ed.\hskip 1em
  plus 0.5em minus 0.4em\relax Iowa State University Press, 1967.

\bibitem{Gonzalez1977}
T.~Gonzalez, S.~Sahni, and W.~R. Franta, ``An efficient algorithm for the
  {K}olmogorov-{S}mirnov and {L}illiefors tests,'' \emph{ACM Transactions on
  Mathematical Software}, vol.~3, no.~1, pp. 60--64, Mar. 1977.

\bibitem{Mintchev:2000}
M.~P. Mintchev, A.~Girard, and K.~L. Bowes, ``Nonlinear adaptive noise
  compensation in electrogastrograms recorded from healthy dogs,'' \emph{IEEE
  Transactions on Biomedical Engineering}, vol.~47, no.~2, pp. 239--248, Jan.
  2000.

\bibitem{Chapa2000}
J.~O. Chapa and R.~M. Rao, ``Algorithms for designing wavelets to match a
  specified signal,'' \emph{IEEE Transactions on Signal Processing}, vol.~48,
  no.~12, pp. 3395--3406, Dec. 2000.

\bibitem{Tew1992}
A.~H. Tewfik, D.~Sinha, and P.~Jorgensen, ``On the optimal choice of a wavelet
  for signal representation,'' \emph{{IEEE} Transactions on Information
  Theory}, vol.~38, no.~2, pp. 747--765, Mar. 1992.

\bibitem{Zhou1993}
H.~Zhou and A.~H. Tewfik, ``Parametrization of compactly supported orthonormal
  wavelets,'' \emph{IEEE Transactions on Signal Processing}, vol.~41, no.~3,
  pp. 1428--1431, Mar. 1993.

\bibitem{Gopi1994}
R.~A. Gopinath, J.~E. Odegard, and C.~S. Burrus, ``Optimal wavelet
  representation of signals and the wavelet sampling theorem,'' \emph{{IEEE}
  Transactions on Circuits and Systems---{II}: Analog and Digital Signal
  Processing}, vol.~41, no.~4, pp. 262--277, Apr. 1994.

\bibitem{SanMiguel:1999}
C.~P. Sanmiguel, M.~P. Mintchev, and K.~L. Bowes, ``Dynamics of level of
  randomness of electrogastrograms can be indicative of gastric electrical
  uncoupling in dogs,'' \emph{Digestive Diseases and Sciences}, vol.~44, no.~3,
  pp. 523--528, Mar. 1999.

\end{thebibliography}
